\documentclass[pdflatex,sn-nature]{sn-jnl}

\usepackage{graphicx}%
\usepackage{multirow}%
\usepackage{amsmath,amssymb,amsfonts}%
\usepackage{amsthm}%
\usepackage{mathrsfs}%
\usepackage[title]{appendix}%
\usepackage{xcolor}%
\usepackage{textcomp}%
\usepackage{manyfoot}%
\usepackage{booktabs}%
\usepackage{algorithm}%
\usepackage{algorithmicx}%
\usepackage{algpseudocode}%
\usepackage{listings}%

\usepackage{graphicx}  
\usepackage{dcolumn}   
\usepackage{bm}        

\usepackage{pdfpages}

\usepackage{physics}
\usepackage{xcolor}
\usepackage[normalem]{ulem}

\theoremstyle{thmstyleone}%
\theoremstyle{thmstyletwo}%

\theoremstyle{thmstylethree}%

\newcommand{\threej}[6]{%
\begin{pmatrix}
#1 & #2 & #3\\
#4 & #5 & #6
\end{pmatrix}}

\newcommand{\cg}[6]{%
\langle #1\,#2\,#3\,#4 | #5\,#6 \rangle}

\begin{document}
	
\title[Article Title]{Finite-Temperature Toroidal Moment Amenable to Direct Observation in an Fe$_{10}$Dy$_{10}$ Molecular Ring}
	

\author*[1]{\fnm{Alessandro}\sur{Soncini}}\email{alessandro.soncini@unipd.it}

\author[2]{\fnm{Kieran}\sur{Hymas}}

\author[3,4,5]{\fnm{Jonas}\sur{Braun}}

\author[3]{\fnm{Yannik F.} \sur{Schneider}}

\author[1]{\fnm{Simone} \sur{Calvello}}

\author[3]{\fnm{Amer} \sur{Baniodeh}}

\author[3]{\fnm{Yanhua} \sur{Lan}}

\author[5,6]{\fnm{Wolfgang} \sur{Wernsdorfer}}

\author[7]{\fnm{Marco} \sur{Affronte}}

\author[3]{\fnm{Christopher E.} \sur{Anson}}

\author*[3,4,5]{\fnm{Annie K.} \sur{Powell}}\email{annie.powell@kit.edu}

\affil*[1]{\orgdiv{Department of Chemical Sciences}, \orgname{University of Padova}, \orgaddress{\street{Via Marzolo 1}, \city{Padova}, \postcode{35123}, \country{Italy}}}

\affil[2]{\orgdiv{Commonwealth Scientific and Industrial Research Organisation}, \orgname{CSIRO}, \orgaddress{\city{Clayton}, \postcode{3168}, \state{Victoria}, \country{Australia}}}

\affil[3]{\orgdiv{Institute of Inorganic Chemistry}, \orgname{Karlsruhe Institute of Technology (KIT)}, \orgaddress{\street{Kaiserstr. 12}, \city{Karlsruhe}, \postcode{ 76131}, \country{Germany}}}

\affil[4]{\orgdiv{Institute of Nanotechnology}, \orgname{Karlsruhe Institute of Technology (KIT)}, \orgaddress{\street{Kaiserstr. 12}, \city{Karlsruhe}, \postcode{ 76131}, \country{Germany}}}

\affil[5]{\orgdiv{Institute for Quantum Materials and Technologies}, \orgname{Karlsruhe Institute of Technology (KIT)}, \orgaddress{\street{Kaiserstr. 12}, \city{Karlsruhe}, \postcode{ 76131}, \country{Germany}}}

\affil[6]{\orgdiv{Physikalisches Institut}, \orgname{Karlsruhe Institute of Technology (KIT)}, \orgaddress{\street{Kaiserstr. 12}, \city{Karlsruhe}, \postcode{ 76131}, \country{Germany}}}

\affil[7]{\orgdiv{Dipartimento di Scienze Fisiche, Informatiche e Matematiche}, \orgname{University of Modena and Reggio Emilia}, \orgaddress{\street{Via Campi 213A}, \city{Modena}, \postcode{ 41125}, \country{Italy}}}

\date{\today}

\abstract{\unboldmath 
		Single-molecule toroics (SMTs) host closed magnetic-vortex configurations that carry toroidal moments $\boldsymbol{\tau}$, whose electric-dipole symmetry enables magnetoelectric spin control. Yet, opposite toroidal chiralities are degenerate in conventional magnetic fields, making direct detection of molecular toroidal polarisation challenging. Current approaches probe molecular toroidal dynamics only indirectly through weak residual magnetism, leaving direct interrogation of toroidal polarisation an open challenge.
        Moreover, the survival of toroidal polarization at finite temperature, and realistic preparation-and-readout conditions, have not been quantitatively established. Here we investigate the icosanuclear $3d$--$4f$ molecular ring Fe$_{10}$Dy$_{10}$, featuring a $\sim$62-billion-dimensional low-energy manifold with pervasive toroidal character, rendered computationally tractable via an ab initio-informed transfer-matrix framework with perturbative corrections. Our model reproduces magnetic and calorimetric measurements and reveals a maximally toroidal ground doublet with robust finite-temperature toroidal response. We introduce the toroidal susceptibility $\xi$ as a finite-temperature linear-response function to quantify toroidal polarisation induced by magnetic-field curl. We then develop a preparation-and-detection protocol in which a temporally asymmetric near-infrared waveform generates a cumulative toroidal population imbalance, while an ab initio-informed magnetoelectric tensor predicts an electric-field-induced magnetic moment within $\mu$SQUID detectability. These results establish Fe$_{10}$Dy$_{10}$ as a molecular platform where toroidal polarisation can be prepared, accumulated and read out under realistic experimental conditions.
}


\maketitle




The interplay between spin–orbit coupling and low-symmetry electrostatic potentials at the nanoscale, such as those present in chiral matter, is emerging as a powerful route to engineer quantum functionalities. It underpins phenomena of direct interest to nanoscience and quantum technologies, including chirality-induced spin selectivity (CISS)\cite{wasielewski2023science}, non-reciprocal dichroism\cite{yokosuk2020nonreciprocal}, magneto-electric coupling~\cite{spaldin2008toroidal,yazback2023toroidal}, and the creation of skyrmionic or antiskyrmionic excitations~\cite{ritzmann2018trochoidal, guan2023optically,bhowal2022magnetoelectric}. These effects exemplify how tailored symmetry breaking at the molecular or nanostructural level can give rise to robust, topologically protected (or dark) quantum states.

Among the most striking manifestations of such symmetry breaking is the toroidal moment, first introduced by Zeldovich to describe electromagnetic multipoles mediating parity-non-conserving weak interactions~\cite{zeldovich1957,dubovik1990}, and later observed in atomic Cs~\cite{wood1997cesium}. This electromagnetic multipole can be pictured as a poloidal current encircling a torus or as a closed loop of magnetic dipoles arranged head-to-tail, resulting in no net magnetic poles—hence the alternative name anapole. Despite its magnetic origin (odd under time reversal), the toroidal moment transforms under spatial inversion like an electric dipole, so that a counter-clockwise magnetic vortex is converted into a clockwise one under parity reversal~\cite{zeldovich1957,dubovik1990,faglioni2004,spaldin2008toroidal}. 
This dual magnetic and electric nature of the toroidal moment implies that a system with toroidal polarisation can support linear magneto-electric coupling~\cite{spaldin2008toroidal}, providing a direct route to spin-electric effects i.e. the electric control of atomic and molecular spins and of their interactions~\cite{ardavan2021natphys,cini2025natcomm,boudalis2025natcomm,vaganov2025natchem,yazback2023toroidal}.  
Moreover, in chiral nanostructures based on SMTs~\cite{zhu2026homochiral}, where parity symmetry is intrinsically broken, uniform magnetic fields can in principle induce toroidal polarisation through mixed magnetic--toroidal response effects, with magnetic and toroidal responses partially superimposed~\cite{faglioni2004}.

For achiral systems, producing a toroidal polarization typically requires inhomogeneous magnetic fields with finite curl $\curl{\bf B}$~\cite{faglioni2004,spaldin2008toroidal}, although recently an EPR-based spectroscopic alternative has been proposed by some of us for non centro-symmetric samples~\cite{hymas2025preparation}.  While classical analogues of toroidal moments have been engineered and detected in non-collinear low-dimensional magnets~\cite{vanaken2007}, and in metamaterials~\cite{zheludev2010,zheludev2022}, to date, microscopic quantum states carrying a molecular toroidal moment have remained experimentally elusive. The reason is mainly two-fold. 

On the one hand, molecular quantum states carrying clockwise and anticlockwise toroidal moments of equal magnitude are necessarily doubly-degenerate in the absence of a finite curl $\curl{\bf B}$,  difficult to produce and control at the scale of a small molecule, which yields a zero net toroidal polarization.  

On the other hand, to date toroidal properties have only been identified and discussed for specific molecular microstates, while the collective role of toroidal excitations within a molecular system and their contribution to a finite-temperature toroidal polarization have largely been ignored, so that a realistic identification of a molecular system where toroidal polarization could be amenable to direct observation at operative temperatures is not currently viable. 
 Both challenges to the observation of a toroidal polarization will be addressed in this study, where we establish a quantitative framework linking finite-temperature toroidal response, dynamical preparation, and experimental readout.
To illustrate this point, we note that the only work in the literature claiming observation of a molecular toroidal state consists of a recent report of a record-long \emph{magnetic} relaxation time detected via $\mu$SQUID magnetometry, attributed to the low-lying ferrotoroidic and antiferrotoroidic states of a Ga$_7$Dy$_6$ Single-Molecule Toroic (SMT)\cite{rajaraman2025ultraslow}. Of course, since the ferrotoroidic states of a centrosymmetric molecule cannot carry a magnetic moment by definition, while an antiferrotoroidic state can be weakly magnetic but cannot carry a net toroidal moment, the involvement of toroidal dynamics may only be inferred indirectly from studying the relaxation of the weakly magnetic non-toroidal states generated by the time-varying uniform magnetic field\cite{rajaraman2025ultraslow}.
	Related observations have also been reported in recent work on nuclear-spin-dependent toroidal ground states in $^{163}$Dy$_4$L$_4$ SMTs~\cite{chen2026toroidal},  as well as in studies employing $^{161}$Dy-M{\"o}ssbauer spectroscopy in a {Co$^{\mathrm{III}}_3$Dy$^{\mathrm{III}}_3$} SMT~\cite{peng2026mossbauer}.
While such inference is interesting, the precise relaxation mechanism of the toroidal states remains silent to typical magnetometry experiments, urging the need for new experimental proposals to detect direct signatures of these elusive molecular states.

Recent studies on transition-metal spin triangles have explored electric-field control of chiral and toroidal-like spin states~\cite{yazback2023toroidal,lewkowitz2023magnetoelectric}. In such weak spin--orbit spin--frustrated systems, toroidal character is closely intertwined with spin chirality and electric-field-induced modifications of spin-Hamiltonian parameters, whereas the direct preparation and readout of a finite-temperature toroidal polarisation in strongly spin--orbit-coupled and non-frustrated SMTs remains an open challenge.
Nevertheless, SMTs remain indeed the most promising candidates where toroidal moments in molecules could in principle be directly observed\cite{tang2006,luzon2008spin,chibotaru2008origin,ungur2012net}, also using their ability to sustain linear magnetoelectric coupling~\cite{popov2009anapole,plokhov2011magnetoelectric}.  They consist of metal ions with strong on-site magnetic anisotropy—typically $4f$ lanthanides—weakly coupled in ring-like topologies that stabilize vortex arrangements of $4f$ electron moments~\cite{soncini2008toroidal, langley2013trinuclear, baniodeh2015ligand, kaemmerer2020inorganic}. SMTs are thus crucial platforms both to probe fundamental molecular physics beyond a simple magnetic dipole picture~\cite{soncini2008toroidal, luzon2008spin, pavlyukh2020toroidal}, and to develop hyper-dense molecular memories~\cite{spaldin2008toroidal}, molecular spintronics devices~\cite{soncini2010toroidalspintronics, pavlyukh2020toroidal} and toroidal qubits~\cite{zagoskin2015toroidalqubit}

Their ground states possess vanishing net magnetic dipole moments but support doubly degenerate, counter-rotating quantum vortices whose toroidal moment is
\begin{equation}\label{eq:1}
	\boldsymbol{\tau}=\sum_i \mathbf{r}_i \times \mathbf{M}_i ,
\end{equation}
where $\mathbf{r}_i$ locates ion $i$ and $\mathbf{M}_i$ is its magnetic moment. Equation (\ref{eq:1}) suggests two chemical design principles for maximizing ground-state toroidal moments: employ ions with large magnetic moments, and enlarge the molecular ring radius. Dysprosium(III) ions, with unquenched orbital momentum and $\sim10\mu_B$ moments, are ideal, and Dy-containing molecular wheels dominate SMT research~\cite{tang2006,luzon2008spin,chibotaru2008origin}. While hexa-, octa- and decanuclear Dy wheels are known~\cite{langley2013trinuclear, tian2015enlarging}, clear evidence of ground-state toroidicity has thus far been limited to Dy$_6$ molecules and only with theoretical support~\cite{ungur2012net, lu2019lanthanide}.

Hybrid architectures introduce additional design freedom. Incorporating $3d$ transition metals to couple Dy$_3$ vortex motifs can generate intramolecular ferrotoroidic states~\cite{vignesh2017ferrotoroidic, vignesh2018slow, kaemmerer2020inorganic, ashtree2021tuning}, while ultra-large heterometallic rings push size and magnetic complexity~\cite{botezat2017ultralarge, baniodeh2018high}. Building on this strategy, some of us reported a family of icosanuclear [Fe$_{10}$Ln$_{10}$] wheels with alternating Fe$^{\text{III}}$ and Ln$^{\text{III}}$ ions in a nanoscale torus~\cite{baniodeh2014unraveling}, where the Gd-analogue showed proximity to a quantum critical point~\cite{baniodeh2018high}. The fixed point in these previous studies was to keep fixed the 3d metal as Fe$^{\text{III}}$, while the effect of varying the Ln$^{\text{III}}$ on magnetic and optical properties was tested. 

Here we investigate the Fe$_{10}$Dy$_{10}$ member of that family, whose large size and complex magnetic structure give rise to a dense manifold of low-lying toroidal microstates. This system provides a unique opportunity to quantify finite-temperature toroidal polarisation as an emergent thermodynamic property arising from multiple thermally accessible toroidal excitations.
Combining multi-reference \emph{ab initio} calculations with a tailor-made perturbatively corrected transfer-matrix approach describing a quantum-decorated non-collinear Ising chain, we quantitatively reproduce the magnetic and calorimetric experimental data and demonstrate an exceptionally large and thermally resilient toroidal response.

We further propose and develop a quantitative framework for the ultrafast dynamical preparation and magnetoelectric readout of molecular toroidal states. In particular, we show that temporally asymmetric near-IR ultrafast electric-field driving can generate an experimentally detectable nonequilibrium toroidal polarisation through the accumulation of population imbalance between time-reversed toroidal states.

Through the development of an \emph{ab initio}-informed model of the magnetoelectric response, we show that the resulting toroidal polarisation generates a linear magnetoelectric effect that can be experimentally probed via solid-state superconducting magnetometers such as $\mu$SQUID techniques.
Analysis of the ultrafast accumulation mechanism further reveals that the accumulated toroidal polarisation scales as $|\boldsymbol{\tau}|^4$, establishing a design principle for large molecular toroidal systems amenable to direct observation.

\section{Results} \label{sec:2}

We investigate the microscopic origin, finite-temperature robustness, and ultrafast preparation and magnetoelectric readout of toroidal states in Fe$_{10}$Dy$_{10}$. Even when restricting the model to the low-energy manifold, the resulting Hilbert space spans $\sim 62$ billion states, beyond the reach of direct {\em ab initio} approaches. We therefore adopt a hybrid atomistic strategy combining {\em ab initio} crystal-field calculations, transfer-matrix methods, nonequilibrium open-system dynamics, and perturbative evaluation of the magnetoelectric response.

\subsection{Ab initio local crystal field states}
 The local states are determined via multiconfigurational scalar relativistic  {\em ab initio} calculations on wheel fragments comprising a single open-shell ion (Dy$^{\text{III}}$ or Fe$^{\text{III}}$) at a time, surrounded by an appropriate ligand environment using the ab initio multiconfigurational software CERES developed by some of us~\cite{calvello2018ceres}.  For all the Dy$^{\text{III}}$ ions, our results indicate energetically well-isolated ground KDs with strongly axial magnetic anisotropy (see Methods~\ref{subsec:cahf}). Interestingly, the Dy$^{\text{III}}$ ions with the more axial g-factors and larger gaps to first excited KD (Dy$_1$, Dy$_2$, Dy$_5$) sit on higher curvature regions of the ellipse.
\begin{figure}[H]
	\centering
	\includegraphics[width=0.45\textwidth]{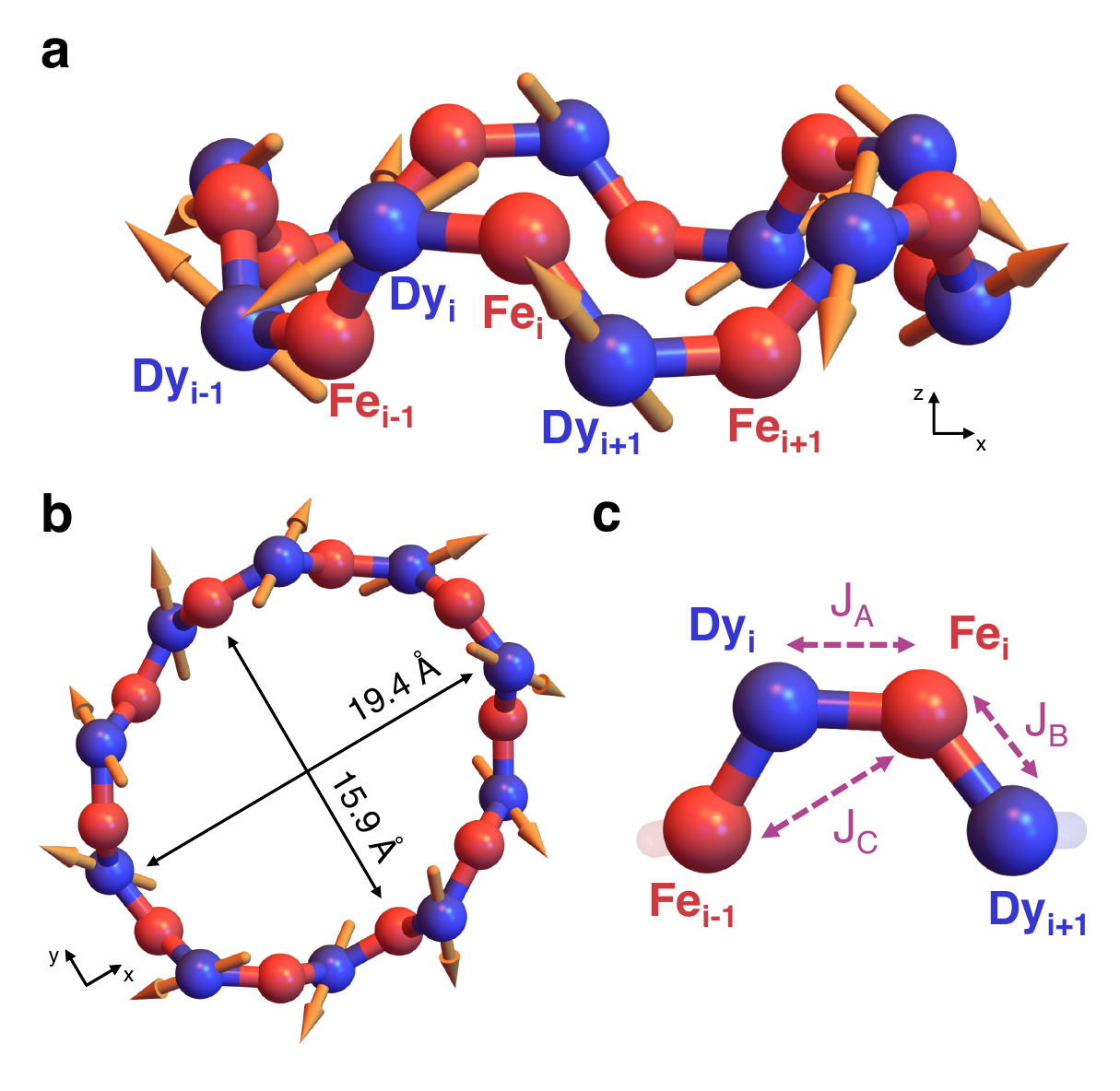}
	\caption{Schematic depiction of the Fe$_{10}$Dy$_{10}$ paramagnetic skeleton. (a) Side profile of the wheel. Fe$^{\text{III}}$ ions are depicted as red spheres and Dy$^{\text{III}}$ ions as blue spheres. The {\em ab initio}-calculated principal magnetic axis $\mathbf{u}_i$ of each of the Dy$^{\text{III}}$ ground doublets (reported in Table \ref{tab:1}) are shown as orange arrows. (b) Birds eye view of Fe$_{10}$Dy$_{10}$ showing the `long' and `short' axes of the ellipse. (c) Exchange connectivity of a repeating fragment of the ring. The Dy-Fe exchange coupling between Dy$^{\text{III}}_i$ and Fe$^{\text{III}}_i$ and between Fe$^{\text{III}}_i$ and Dy$^{\text{III}}_{i+1}$ are $J_A$ and $J_B$, respectively. The weaker Fe$^{\text{III}}$-Fe$^{\text{III}}$ antiferromagnetic exchange coupling is $J_C$.}
	\label{fig:1}
\end{figure}
The magnetic axes are not perfectly planar with respect to the median plane of the Fe$_{10}$Dy$_{10}$ wheel  but instead observe an alternating canting pattern about the ring (Figure~\ref*{fig:1}a), consistent with the molecular symmetry. From Figure \ref{fig:1}b it is nonetheless clear that the planar projections of the axes describe a vortex like arrangement which is a necessary (though not a sufficient) condition for the existence of a ground state toroidal moment. Previous calculations on a related Fe$_{8}$Dy$_{8}$ ring were reported, where perhaps surprisingly the orientations of the Dy Ising axes are not consistent with the molecular S$_{8}$ symmetry.  Further calculations for each of the symmetry unique Fe$^{\text{III}}$ ions using MOLCAS~\cite{aquilante2016molcas} revealed energetically well-isolated orbitally non-degenerate $S^{\text{Fe}}=5/2$ pure spin ground multiplets with negligible zero-field splitting.

\subsection{DFT magnetic coupling constants}

Exchange coupling in Fe$_{10}$Dy$_{10}$ was modelled via broken symmetry density functional theory (DFT) calculations for each Fe$^{\text{III}}$-Dy$^{\text{III}}$-Fe$^{\text{III}}$ fragment (see Methods~\ref{subsec:DFT} ).We utilised Yamaguchi's method~\cite{yamaguchi1986molecular} to extract Heisenberg exchange coupling constants $J_A$ and $J_B$ between the Fe$^{\text{III}}$ and neighbouring Dy$^{\text{III}}$ spins as well as a next-nearest neighbour constant $J_C$ to account for Fe$^{\text{III}}$-Fe$^{\text{III}}$ exchange.   A schematic of the exchange connectivity is given in Figure \ref{fig:1}c while the calculated constants are reported in Table \ref{tab:2}. 
\begin{table}[h]
	\caption{\label{tab:2}
		Exchange coupling constants determined via broken symmetry density functional theory calculations for symmetry unique fragments of Fe$_{10}$Dy$_{10}$. $J_A$ and $J_B$ are nearest-neighbour Fe$^{\text{III}}$-Dy$^{\text{III}}$ and $J_C$ are next-nearest neighbour Fe$^{\text{III}}$-Fe$^{\text{III}}$ exchange couplings as depicted in Figure~\ref{fig:1}c.}
	\begin{tabular}{cccc}
		\toprule
		\textrm{Fragment}&
		\textrm{$J_A/$cm$^{-1}$}&
		\textrm{$J_B/$cm$^{-1}$}&
		\textrm{$J_C/$cm$^{-1}$}\\
		\midrule
		Fe$_{10}$-Dy$_{1}$-Fe$_{1}$ & 0.3267 & 1.041 & -0.1682\\
		Fe$_{1}$-Dy$_{2}$-Fe$_{2}$ & 0.2854 & 1.125 & -0.1656\\
		Fe$_{2}$-Dy$_{3}$-Fe$_{3}$ & 0.5771 & 0.998 & -0.1657\\
		Fe$_{3}$-Dy$_{4}$-Fe$_{4}$ & 0.3614 & 0.862 & -0.2597\\
		Fe$_{4}$-Dy$_{5}$-Fe$_{5}$ & 0.1325 & 1.028 & -0.0263\\
		\botrule
	\end{tabular}
\end{table}

Our calculations revealed asymmetric ferromagnetic exchange couplings between Fe$^{\text{III}}$ and Dy$^{\text{III}}$ nearest neighbours, with Dy ions on lower curvature regions (Dy3, Dy4) associated with stronger $J_A$ and weaker $J_B$, and a systematically weaker antiferromagnetic coupling between Fe$^{\text{III}}$-Fe$^{\text{III}}$ next-nearest neighbours.  Theoretical modelling of an analogous Fe$_{10}$Gd$_{10}$ system~\cite{baniodeh2018high} demonstrated a similar pattern.

To further assess the robustness of these exchange parameters, we performed an independent statistical $\chi^2$ analysis based on a three-parameter fit of the magnetic susceptibility data (Supplementary Note 8). The best-fit values, $J_A^{\mathrm{fit}}=0.8^{+0.2}_{-0.55}$ cm$^{-1}$, $J_B^{\mathrm{fit}}=0.6^{+0.7}_{-0.15}$ cm$^{-1}$ and $J_C^{\mathrm{fit}}=-0.10^{+0.10}_{-0.05}$ cm$^{-1}$, are consistent with the broken-symmetry DFT estimates, which lie within the 68\% confidence region of the fit. This independent analysis validates both the magnitude and sign of the dominant exchange interactions used in the ab initio-parameterised transfer-matrix model.

\subsection{Eigenstates and finite-temperature response properties}

The magnetic excitations of Fe$_{10}$Dy$_{10}$  were obtained via the following spin Hamiltonian, whose parameters are evaluated via ab initio calculations:
\begin{equation} \label{eq:2}
	H = H_{\text{ex}} + H_{\text{dip}} + H_{\text{Zee}} + H_{\text{Tor}} ,
\end{equation}
where $H_{\text{ex}}$ and $H_{\text{dip}}$ are the isotropic exchange and dipolar interaction terms, respectively (see Eqs.~(\ref{eq:3}, \ref{eq:4}) in Methods), $H_{\text{Zee}}$ is the Zeeman Hamiltonian describing the coupling to a uniform external magnetic field (Eq.~\ref{eq:5} in Methods), and  $H_{\text{Tor}}$ is an extension of the Zeeman Hamiltonian (Eq.~\ref{eq:6} in Methods) accounting for the coupling of the wheel's toroidal moment to $\nabla\times\mathbf{B}$ ({\em vide infra}). From the ab initio results, it is clear that for $T< 90K$  Eq.~(\ref{eq:2}) can be projected on the product basis $\left| \bm{\sigma}\bm{M}\right\rangle$, where the $2^{10}$ Ising configurations are labeled by $\boldsymbol{\sigma} = \left( \sigma_1, \dots, \sigma_{10} \right)$, with $\sigma_i = \pm 1$ selecting between the two components of the ground state easy-axis KDs of the $i^{\text{th}}$ Dy$^{\text{III}}$ ion, consisting of almost pure $M_J=\pm15/2$ total angular momentum states projected on their respective non-collinear principal magnetic axes $\mathbf{u}_i$ (see Table \ref{tab:1}), while the $6^{10}$ spin product states are labeled by$\bm{M} = {M_1, M_{2}, \dots M_{10}}$, with  $M_i = -5/2, -3/2, \dots +5/2$ are the degenerate spin states of the $S=5/2$ centered on the $i^{\text{th}}$ Fe$^{\text{III}}$ ion. 
In this approximation, the ring Hamiltonian is block diagonal with respect to the $2^{10}$ possible Ising configurations, and so each block $\boldsymbol{\sigma}$ may be considered individually (see Methods~\ref{subsec:ham}).
 
Note that Eq.~(\ref{eq:2}), while block-diagonal on the Ising $\bm{\sigma}$ configurations, would still give rise to $2^{10}$ diagonalisation problems with dimension $6^{10}\simeq 60.5$ million, which is too computationally demanding.  However, ab initio results show that Fe$^{\text{III}}$-Fe$^{\text{III}}$  exchange is significantly weaker than Dy$^{\text{III}}$-Fe$^{\text{III}}$ coupling. Thus we adopted the perturbative strategy described in Methods~\ref{subsec:fefept} to account for the spin-frustrating contribution of Fe$^{\text{III}}$-Fe$^{\text{III}}$ exchange, which keeps the problem non-diagonal only on the local spin basis of Fe$^{\text{III}}$ quantum decoration, and solely dependent on the specific $\boldsymbol{\sigma}$.

Finally, the free energy $F$ and the ensuing magnetic response properties are computed via a transfer matrix approach, as detailed in Methods~\ref{subsec:transfermat} .

\subsection{Low energy spectrum}\label{subsec:spectrum}

\begin{figure*}[h]
	\centering
	\includegraphics[width=0.95\textwidth]{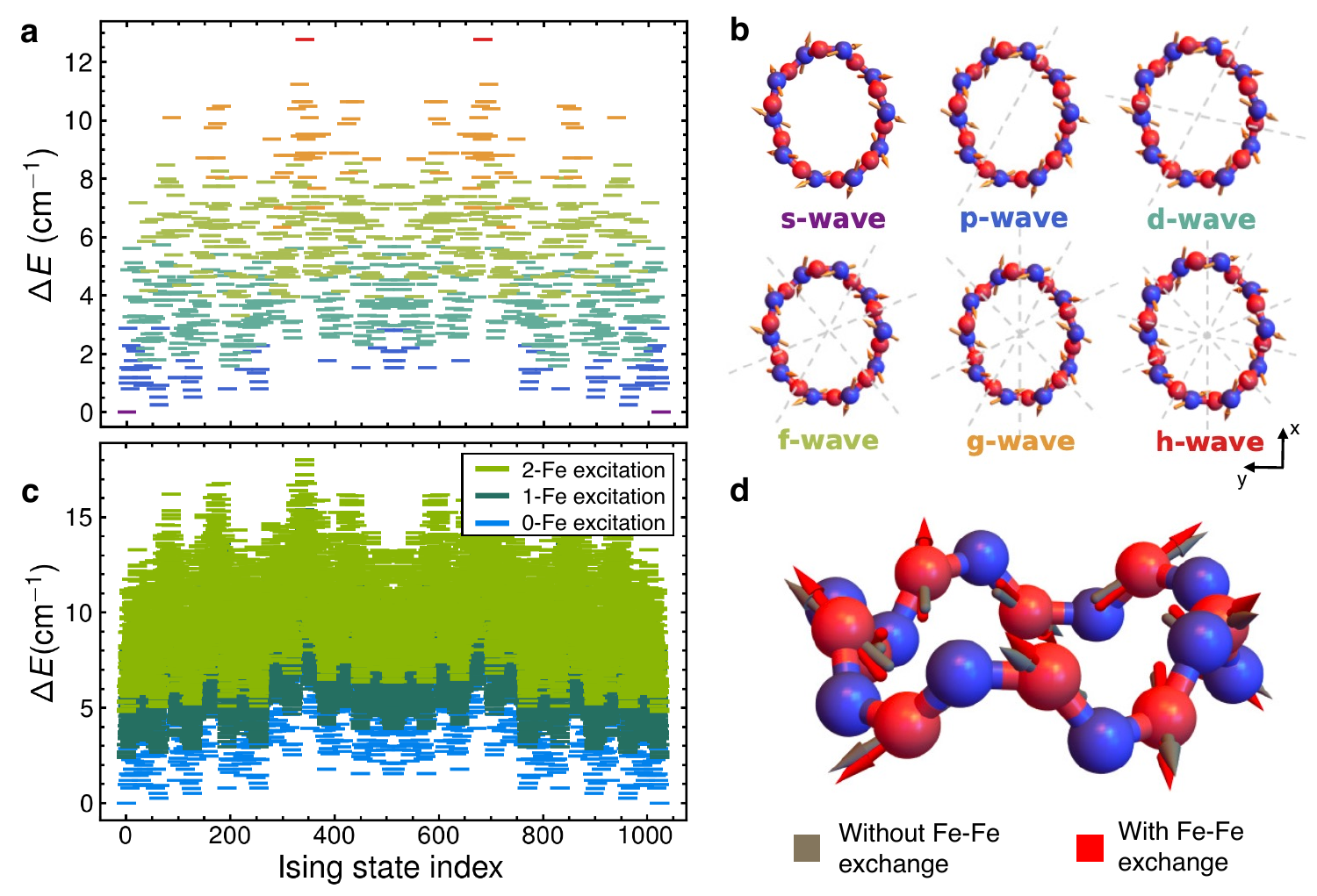}
	\caption{Low energy spectrum of Fe$_{10}$Dy$_{10}$.(a) Lowest energy (b) Archetypal examples of s, p, d, f, g and h-wave Ising spin configurations analogous to the spin wave nomenclature. Colour coding corresponds to the band structure shown in (a). (c) Ground and excited state band structure of Fe$_{10}$Dy$_{10}$ from all possible one and two Fe$^{\text{III}}$ excitations around the ring for fixed Dy$^{\text{III}}$ Ising spin configurations. The Ising spin configurations are indexed by mapping the vector $\boldsymbol{\sigma}$ to a binary string and straightforwardly converting to a decimal representation offset by $1$. (d) Calculated ground state Fe$^{\text{III}}$ on-site spin expectation values for the toroidal Dy$^{\text{III}}$ Ising spin configuration calculated with (red arrows) and without (grey arrows) Fe$^{\text{III}}$-Fe$^{\text{III}}$ antiferromagnetic exchange. For the calculations involving Fe$^{\text{III}}$-Fe$^{\text{III}}$ exchange, first order perturbation theory corrections to the wavefunction were considered taking corrections from 1-Fe exchange.}
	\label{fig:5}
\end{figure*}
Using Eq. (\ref{eq:9}, \ref{eq:10}) we plotted the spectrum of the $2^{10}$ Dy$^{\text{III}}$ Ising spin configurations $\boldsymbol{\sigma}$ (unflipped Fe spins). In Figure \ref{fig:5}a we highlight classes of Ising excitations according the number of angular nodes distorting the zero-noded vortex pattern (s-wave) of the Dy magnetic moments, analogously to angular spin-wave excitations~\cite{akhiezer1968spinwaves} (explicit representations in Figure \ref{fig:5}b). Note that most of the non-h-wave excitations are endowed with both toroidal and magnetic moments of variable sizes (see Supplementary Note 2).
The modest ellipticity of the Fe$_{10}$Dy$_{10}$ ring (major/minor axis ratio $\sim1.22$) may contribute to reducing the energy gap ($\sim0.26$ cm$^{-1}$) between the toroidal ground state ($s$-wave) and the competing onion-like ($p$-wave) magnetic excitations, consistent with previous studies of vortex states in elliptical magnetic rings.~\cite{gaididei2019}
In Figure \ref{fig:5}c we also show the stacks of all Fe-spin single and double excitations.

Similarly to Eq. (\ref{eq:10}), we derived perturbative corrections to the on-site ground state spin expectation values of the Fe$^{\text{III}}$ ions considering the next-nearest neighbour Fe$^{\text{III}}$-Fe$^{\text{III}}$ exchange in first order perturbation theory, including only single excitations to correct the local Fe ground-state wavefunctions, a choice supported by Figure \ref{fig:5}c.  In Figure \ref{fig:5}d we plot the corrected average spin values of the Fe$^{\text{III}}$ ions (red) superimposed to their unperturbed values (grey). Unsurprisingly, the antiferromagnetic Fe-Fe exchange coupling directs the Fe spin moments towards a N\'eel configuration which tends to diminish the ring toroidal moment~\cite{baniodeh2014unraveling}. 

\subsection{Comparison with experiments}\label{sec:ComparisonExperiments}

Using a Quantum Design SQUID magnetometer MPMS-XL, we performed isothermal powder magnetisation and variable-temperature ($1.8$--$300$ K) magnetic susceptibility measurements on Fe$_{10}$Dy$_{10}$. Further magnetic data are provided in the Supplementary Information. From our transfer-matrix model, the magnetisation ($M_\alpha$) and magnetic susceptibility ($\chi_{\alpha\beta}$), together with their powder averages, are readily calculated from Eq.~(\ref{eq:14}) via derivatives of the free energy with respect to the external magnetic field $\mathbf{B}$:
\begin{equation}\label{eq:15}
M_\alpha=-\frac{\partial F}{\partial B_\alpha},
\qquad
\chi_{\alpha\beta}=-\frac{\partial^2F}{\partial B_\alpha\partial B_\beta}.
\end{equation}
To account for thermal population of excited Dy$^{\rm III}$ doublets when simulating $\chi_{\alpha\beta}$, we employed $\chi_m T = (\chi_{xx}T+\chi_{yy}T+\chi_{zz}T)/3 + (\chi T)_{\rm ab\ initio} - (\chi T)_0$, where the first term is the low-temperature powder average from Eq.~(\ref{eq:15}), $(\chi T)_{\rm ab\ initio}$ is the sum of single-ion {\em ab initio} $\chi T$ contributions, and $(\chi T)_0$ avoids double counting of the Dy$^{\rm III}$ ground doublets and Fe$^{\rm III}$ ground spin manifold~\cite{vignesh2017ferrotoroidic}.

\begin{figure}[H]
	\centering
	\includegraphics[width=0.47\textwidth]{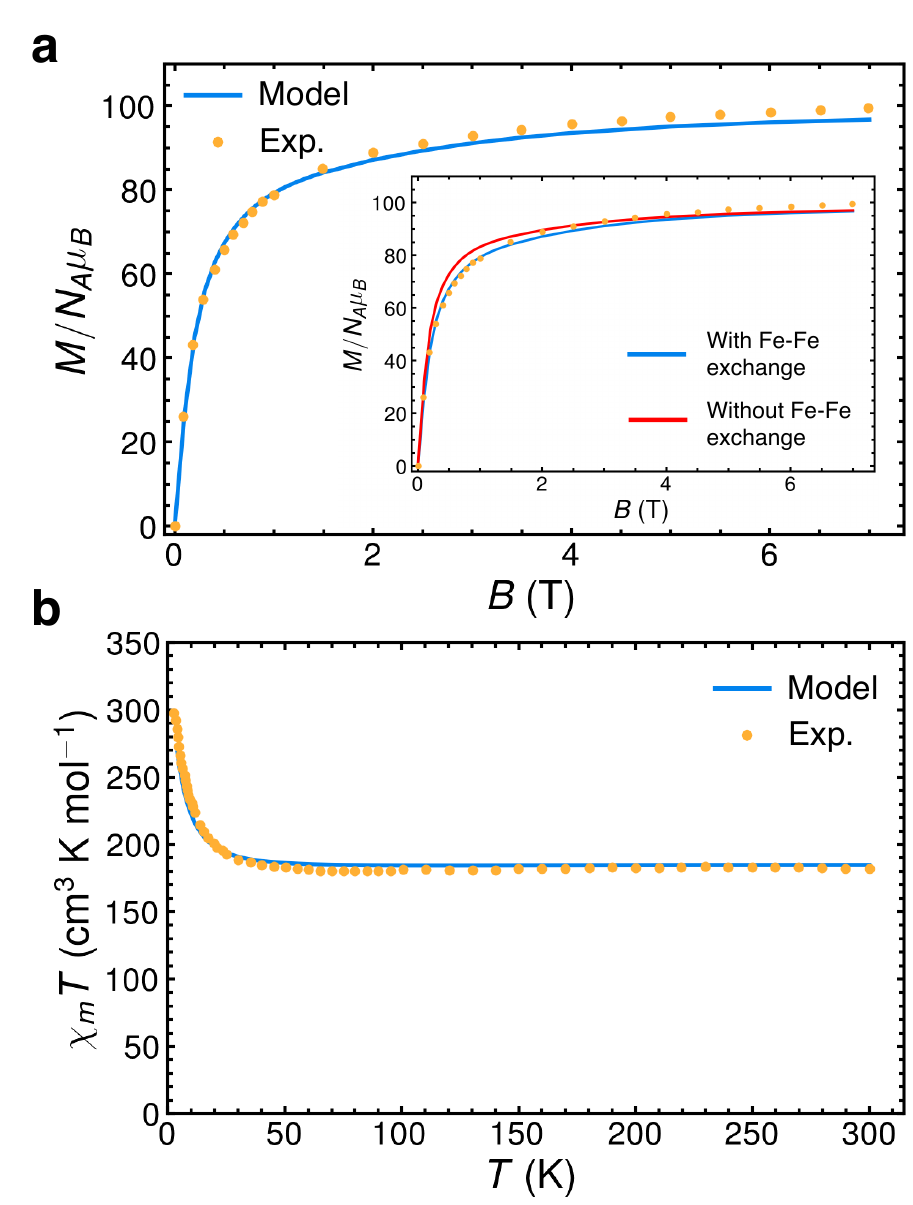}
	\caption{Magnetic measurements and simulated properties of Fe$_{10}$Dy$_{10}$. (a) Comparison between experimental measurements (orange circles) and simulated (solid blue line) isothermal powder magnetisation at $T = 2$ K. Inset: comparison between the experimental and simulated $T = 2$ K powder magnetisation when Fe$^{\text{III}}$-Fe$^{\text{III}}$ next-nearest neighbour exchange $J_C$ is switched on/off (blue/red curves) in our model. (b) Comparison between experimental (orange circles) and simulated (solid blue line) molar magnetic susceptibility $\chi_m T$ as a function of temperature.}
	\label{fig:2}
\end{figure}

\begin{figure}[H]
	\centering
	\includegraphics[width=0.45\textwidth]{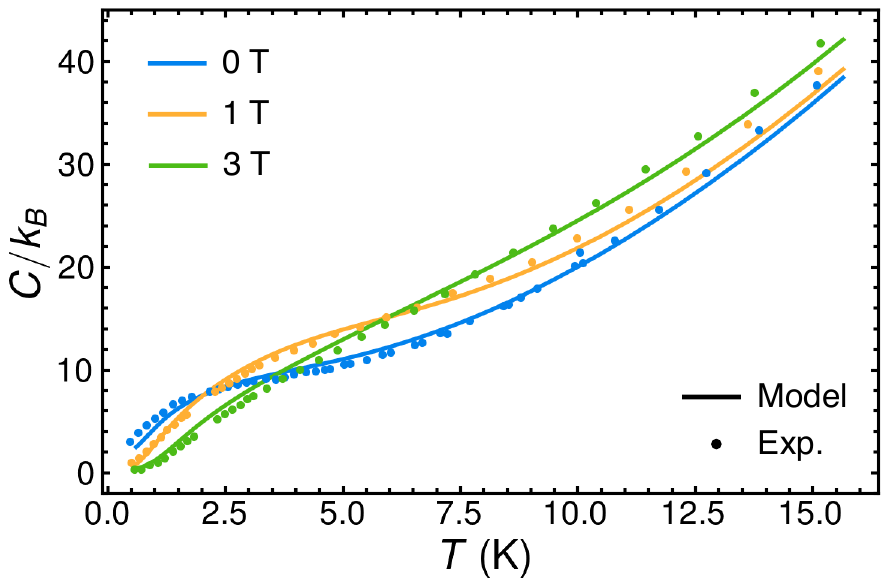}
	\caption{Specific heat of Fe$_{10}$Dy$_{10}$. Specific heat measurements (circles) on a Fe$_{10}$Dy$_{10}$ powder sample as a function of temperature for several values of applied magnetic field $\abs{\mathbf{B}} = 0$ T (blue), $1$ T (yellow) and $3$ T (green). Simulations (solid lines) using our transfer matrix mode averaged over angular distribution.}
	\label{fig:3}
\end{figure}

In Figure \ref{fig:2} we report excellent agreement between magnetic measurements and our {\em ab initio}-parameterised simulations.  Notably, magnetic data seem to indicate a ground state with a large magnetic moment. However, our simulations show that a state with magnetic moment $\sim 83 \ \mu_B$ lies $\sim0.26$cm$^{-1}$ above the magnetically compensated toroidal ground doublet, so that a field of $\sim$6mT along the easy axis is sufficient to make the ground state magnetic. Towards higher fields, our simulations undershoot the experiment slightly which could be a sign of magnetic torquing in the experiment not accounted for in our simulations. To demonstrate the spin-frustrating effect of Fe$^{\text{III}}$-Fe$^{\text{III}}$ exchange coupling, in the inset of Figure \ref{fig:2}a we present simulations of the $T=2$ K powder magnetisation with (solid blue curve) and without (solid red curve) Fe$^{\text{III}}$-Fe$^{\text{III}}$ coupling, unequivocally demonstrating that our perturbative inclusion of Fe$^{\text{III}}$-Fe$^{\text{III}}$  exchange is crucial to reproduce the low field ($\abs{\mathbf{B}} \leq 3$ T) powder magnetisation.
In addition to magnetic measurements, we also collected specific heat data for Fe$_{10}$Dy$_{10}$ using a commercial PPMS ${}^3$He system from Quantum Design. Heat capacity was measured on pressed pellets of micro-crystals with approximate weight $1$-$2$ mg by using the two-tau relaxation method. In Figure \ref{fig:3} we present the data collected at a range of applied magnetic fields (coloured circles) as well as theoretical simulations using our transfer matrix model (solid lines).
Once again our simulations, using the formula~\cite{affronte2000specificheatfitting}
\begin{equation} \label{eq:17}
C/k_B = -T \frac{\partial^2 F}{\partial T^2} + a T^3 + b T^{3/2},
\end{equation}
provide excellent agreement with the experiments.  Importantly, we capture the shifting of the Schottky barrier to higher temperatures with the application of external magnetic field. In addition to the specific heat from our transfer matrix model, we included the terms $aT^3$ and $b T^{3/2}$ in Eq. (\ref{eq:17}) accounting phenomenologically for lattice contributions to the specific heat.  The first contribution is the well known Debye term, while the second accounts for anharmonic corrections which were necessary in order to reproduce the high temperature data~\cite{affronte2000specificheatfitting}, and are here described by the only two fitting parameters $a = 1 \times 10^{-3} \ k_B^{-1}$ K$^{-3}$, and $b = 1 \times 10^{-3} \ k_B^{-1}$ K$^{-3/2}$.

\subsection{Finite-temperature molecular toroidal response} \label{sec:MaxTorState}

\begin{figure}[H]
	\centering
	\includegraphics[width=0.45\textwidth]{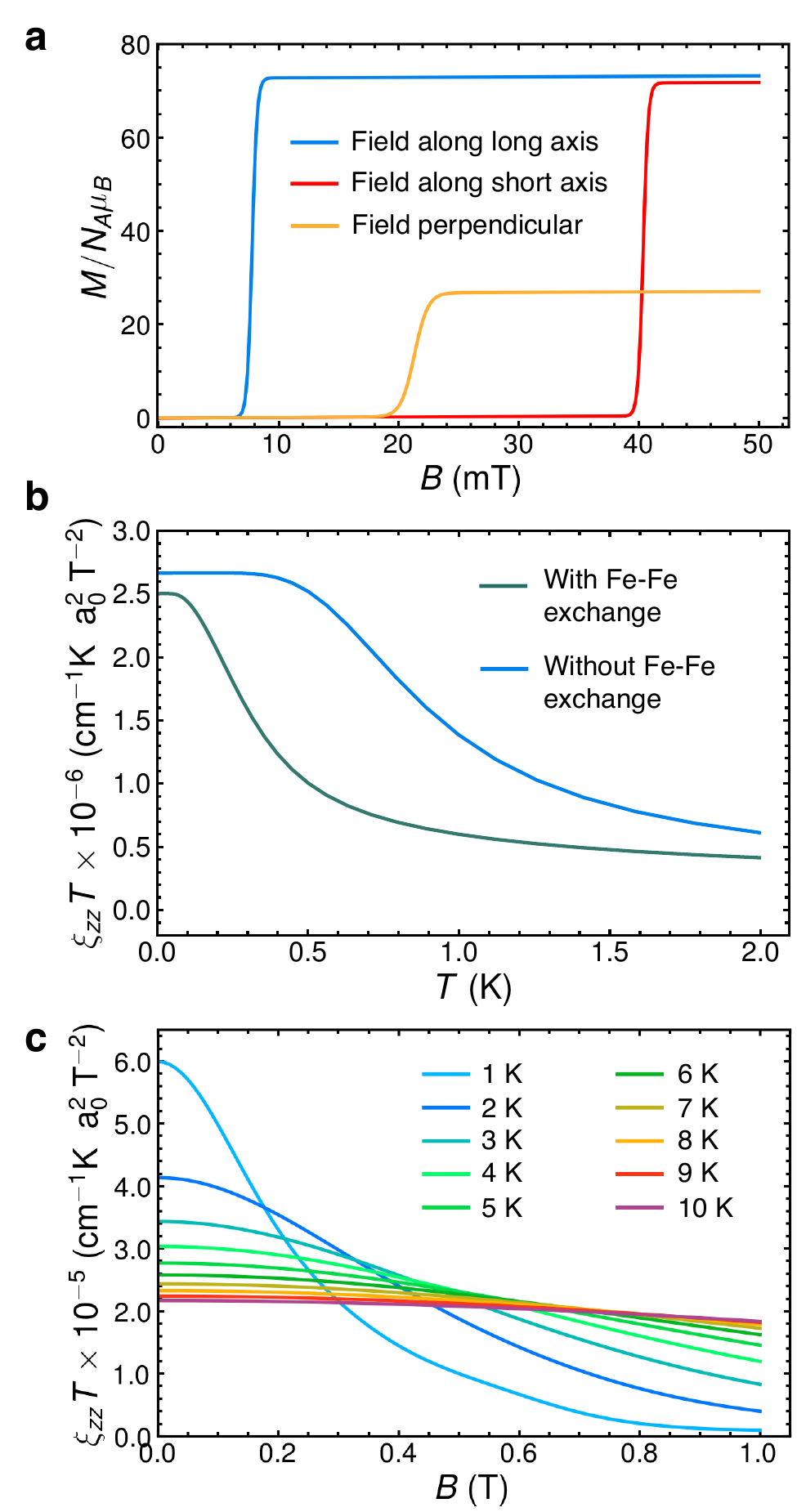}
	\caption{Evidence for a toroidal ground state. (a) $T=0.01$ K single-crystal magnetisation simulations using our {\em ab initio}-parameterised transfer matrix model. The field is oriented along the long axis (blue curve), the short axis (red curve) and perpendicular to the Fe$_{10}$Dy$_{10}$ ring (yellow curve). (b) Low temperature zero-curl toroidal susceptibility simulations with (blue) and without (red) Fe$^{\text{III}}$-Fe$^{\text{III}}$ exchange coupling included in the model. (c) Dependence on an applied uniform magnetic field of the toroidal susceptibility, calculated at various temperatures ranging from 1K to 10K.}
	\label{fig:4}
\end{figure}

While the magnetic measurements at $T= 2$ K indicate sizable low-temperature magnetic response of Fe$_{10}$Dy$_{10}$, they do not provide a characterisation of the zero-temperature ground state, nor do they exclude the existence of finite-temperature toroidal properties.  In Figure \ref{fig:4}a we simulate the static single-crystal magnetisation of Fe$_{10}$Dy$_{10}$ at $T= 0.01$ K, consistently showing a sharp rise in the magnetisation at a finite value of the field, suggesting a magnetically compensated ground state of Fe$_{10}$Dy$_{10}$.
To probe the toroidal properties of the SMT at finite temperature, we introduce the toroidal susceptibility tensor $\xi_{\alpha \beta}$, describing the linear response of the system to a magnetic-field curl $\nabla \times \mathbf{B}$. While quantities related to finite-temperature toroidal response have previously appeared in Landau expansions of ordered states~\cite{planes2014thermotoroidal}, or in calculations of spin toroidization in periodic systems beyond the linear-response regime~\cite{gao2018spintoroidizationl}, the present $\xi_{\alpha \beta}$ is defined as an explicit finite-temperature linear-response tensor, directly analogous to the Van Vleck magnetic susceptibility, and specifically formulated for \emph{molecular} systems with discrete, thermally populated toroidal spectra:
\begin{equation} \label{eq:18}
	\xi_{\alpha \beta} = - \left. \frac{\partial^2 F}{\partial \left( \nabla \times \mathbf{B} \right)_{\alpha} \partial \left( \nabla \times \mathbf{B} \right)_{\beta}} \right|_{\nabla \times \mathbf{B} = 0}.
\end{equation}
Considering the free energy expressed in cm$^{-1}$ and the magnetic-field curl in T $a_0^{-1}$ ($a_0$ the Bohr radius), $\xi_{\alpha\beta}$ is reported in units of cm$^{-1} a_0^2$ T$^{-2}$. 
In direct analogy with the Van Vleck magnetic susceptibility, $\xi_{\alpha \beta}$ can be expressed as an analytical derivative of the free energy~\cite{vansonc2013derivahelmholtz}:
\begin{equation}
\begin{aligned}\label{xianalytical}
\xi_{\alpha \beta} &= \frac{1}{Z} \sum\limits_{n} e^{-\beta E_n} \left[ \beta \sum\limits_{\nu \mu} \bra{n \nu} \tau_{\alpha} \ket{n \mu} \bra{n \mu} \tau_{\beta} \ket{n \nu} \right.\\
& \qquad \qquad + \left. \sum\limits_{m \neq n} \sum\limits_{\nu \mu} \frac{\bra{n \nu} \tau_{\alpha} \ket{n \mu} \bra{n \mu} \tau_{\beta} \ket{n \nu}}{E_m - E_n} \right]
\end{aligned}
\end{equation}
involving matrix elements of the toroidal moment operator $\tau_{\alpha}$ defined in Eq. (\ref{eq:1}).   
Here all toroidal matrix elements are evaluated with respect to the molecular inertia frame. In general, toroidal moments are strictly origin independent only for magnetically compensated states.\cite{Ederer2007,faglioni2004,spaldin2008toroidal} In this system, the maximally toroidal ground-state is thus rigorously origin independent, whereas finite-temperature corrections to $\xi_{\alpha\beta}$ arising from thermally populated excited states can inherit a corresponding origin dependence. The molecular inertia frame nevertheless provides a useful reference frame both for analysing single-molecule toroidal structure-property correlations and for constructing a compact multipolar energy expansion of the linear response to locally constant magnetic-field curls in single-molecule settings, or to globally constant magnetic-field curls in molecular ensembles.

Eq.~(\ref{xianalytical}) therefore describes the finite-temperature molecular toroidal response of an ensemble of non-interacting SMTs in the toroidal analogue of a conventional paramagnetic regime, rather than a spontaneously ordered ferrotoroidic phase. In zero magnetic-field curl the equilibrium ensemble remains globally unpolarised, while the induced linear toroidal polarisation generated by an external curl perturbation is

\begin{equation}\label{taulinear}
\tau_\alpha(T) = \sum_\beta \xi_{\alpha\beta}(T)\left(\nabla \times \mathbf{B}\right)_\beta.
\end{equation}

We plotted the $\xi_{\alpha \beta}$ component along the ring axis as function of temperature in Figure \ref{fig:4}b with and without Fe$^{\text{III}}$-Fe$^{\text{III}}$ exchange. In the limit $T \rightarrow 0$, $\xi_{\alpha \alpha} T = \abs{\bra{0} \tau_{\alpha} \ket{0}}^2/k_B$, hence the low-temperature value $\xi_{zz}T \sim 2.5 \times 10^6$ cm$^{-1}$ K $a_0^2$ T$^{-2}$ yields a ground-state toroidal moment $\abs{\bra{0} \tau_{\alpha} \ket{0}} \sim 1.3 \times 10^3$ $a_0\mu_B$.

Using the peak magnetic-field curl estimated for the ultrafast near-IR driving protocol discussed in Section~1.7, $(\nabla\times\mathbf B)_z \sim 5\times10^{-4}$ T nm$^{-1} \sim 2.65\times10^{-5}\ {\rm T\,a_0^{-1}}$, Eq.~(\ref{taulinear}) predicts induced toroidal moments $\tau_\alpha(T)\sim 6.1\times10^2$, $\sim 5.3\times10^1$, and $\sim 1.5\times10^1$ $a_0\mu_B$ at $T=0.1$, $0.5$, and $1$ K, respectively, showing that even at $0.1$ K the induced finite-temperature toroidal polarisation remains about $45\%$ of the fully saturated ground-state toroidal moment, before progressively degrading at higher temperatures.
We show in Figure \ref{fig:4}b that in both regimes of Fe$^{\text{III}}$-Fe$^{\text{III}}$ exchange, a sizable ground state toroidal moment can be expected at zero temperature, although antiferromagnetic Fe$^{\text{III}}$-Fe$^{\text{III}}$ exchange reduces its magnitude due to spin-frustrating effects in toroidal rings with $N > 3$ centres~\cite{soncini2008toroidal}. Also, we note that in the absence of Fe$^{\text{III}}$-Fe$^{\text{III}}$ exchange coupling $\xi_{\alpha \beta}$ has a slower decay with temperature, suggesting that Fe$^{\text{III}}$-Fe$^{\text{III}}$ exchange lowers in energy magnetic states becoming thermally populated at lower temperatures.  However, as shown in Figure \ref{fig:4}c, we also find that for the same reason, applying a magnetic field will uniformly spread a large number of toroidal states across the spectrum, so that up to $T\sim$10K the magnetic field stabilizes a constant toroidal polarization that at finite temperatures is larger than what is achievable at zero field.

\par\addvspace{1.0\baselineskip}


\subsection{Ultrafast-driven generation and magnetoelectric readout of toroidal polarisation}\label{sec:ultrafast}

While the equilibrium toroidal susceptibility $\xi$ demonstrates that large finite-temperature toroidal moments are thermodynamically accessible, direct observation requires lifting the degeneracy between opposite toroidal chiralities. We show here that an ultrafast temporally asymmetric electric-field waveform can generate a cumulative toroidal population imbalance detectable via the magnetoelectric response of the polarised sample.

In the absence of free currents, the magnetic-field curl associated with a temporally asymmetric electric-field waveform is governed by the Amp\`ere--Maxwell relation~\cite{jackson2021classical}
\begin{equation}
	\curl{\mathbf{B}(t)} = \mu_0 \epsilon_0 \frac{\partial \mathbf{E}(t)}{\partial t}.
\end{equation}
Although $\partial \mathbf{E}(t)/\partial t$ integrates to zero over each optical cycle, a temporally asymmetric waveform produces a strongly asymmetric distribution of $\partial \mathbf{E}/\partial t$. In the presence of finite relaxation times, this prevents complete cancellation of the toroidal response within each cycle, leading to cumulative toroidal polarisation.

To clarify the physical origin of this mechanism, we first analyse an effective two-level open quantum system within a Lindblad framework (Methods~\ref{subsec:tauviadEdt} and Supplementary Note 7.1), where opposite toroidal chiralities $\pm \boldsymbol{\tau}$ are coupled to the driving field via $H_{\mathrm{drive}} \propto \boldsymbol{\tau} \cdot (\nabla \times \mathbf{B}(t))$ and to a phonon bath characterised by a spin-phonon rate constant $\kappa$~\cite{Breuer2002}. Treating dissipation perturbatively yields a closed-form expression for the toroidal population imbalance accumulated over one optical period $T_l$, which to leading order in $\kappa$ becomes
\begin{equation}\label{tau4}
	\begin{aligned}
		\langle \tau (t) \rangle 
		= \frac{16 \kappa \abs{\boldsymbol{\tau}}^4}{T_l^3} \left( \frac{E_0}{c^2} \right)^3 \frac{\epsilon^3(\epsilon - 4)}{(\epsilon - 2)^2} t + \mathcal{O}(\kappa^2),
	\end{aligned}
\end{equation}
where $\epsilon$ controls the temporal asymmetry of the waveform, $E_0$ is the electric field amplitude, and $c$ is the speed of light.

Although the expansion is formally carried out in powers of $\kappa$, the perturbative coefficients grow with irradiation time through repeated integration of the dissipative transition rates. Consequently, the linear accumulation regime described by Eq.~(\ref{tau4}) is transient and higher-order relaxation processes eventually become important. However, explicit numerical integration of the dissipative many-state dynamics under repeated pulse--wait sequences shows that, for realistic relaxation parameters, the linear regime remains dominant over experimentally relevant timescales.

Equation~(\ref{tau4}) highlights three key features: (i) the accumulated toroidal polarisation initially grows linearly with time under repeated driving cycles, (ii) it depends critically on the asymmetry parameter $\epsilon$, vanishing in the symmetric limit $\epsilon = 4$, and (iii) within the linear regime it scales as $|\boldsymbol{\tau}|^4$, providing a strong nonlinear amplification mechanism favouring systems with large intrinsic toroidal moments such as Fe$_{10}$Dy$_{10}$ over smaller toroics such as Dy$_3$.

While this analytical model establishes the mechanism and scaling laws, quantitative predictions require the full low-energy spectrum and relaxation pathways of Fe$_{10}$Dy$_{10}$. We therefore perform numerical simulations based on the 144 low-energy {\em ab initio}-derived collective states, combined with a Lindblad master-equation approach (Methods~\ref{subsec:tauviadEdt}; Supplementary Note 7.2) accounting for multi-channel relaxation beyond the two-level approximation.

\begin{figure}[t]
			\centering
            \includegraphics[width=0.99\textwidth]{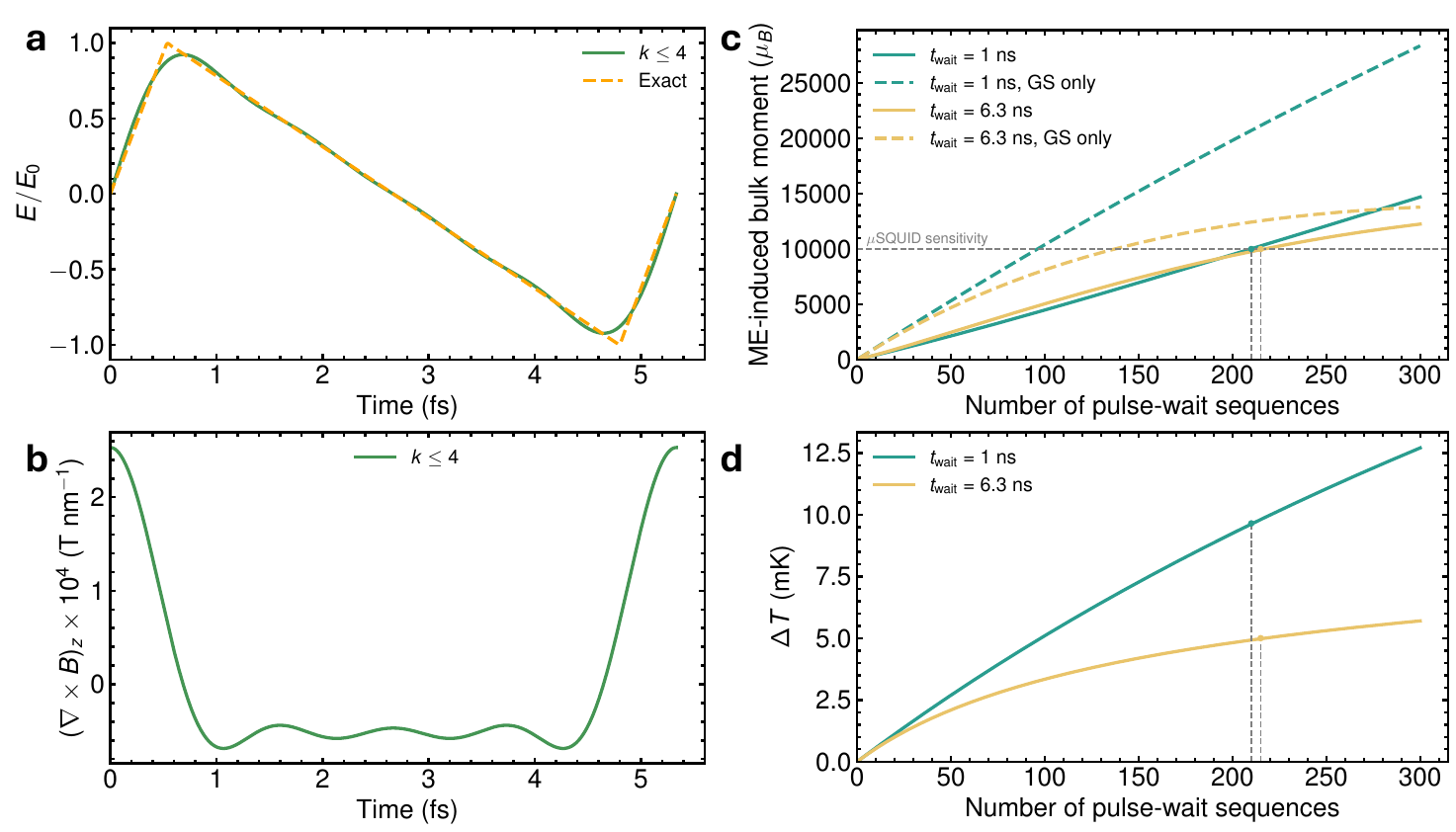}
            \caption{ \textbf{(a)} Temporally asymmetric optical waveform used to drive toroidal dynamics, obtained from the superposition of the first four harmonics of a near-infrared laser field ($\lambda = 1600$ nm), compared with the ideal sawtooth profile.  \textbf{(b)} Corresponding time-dependent magnetic-field curl $(\nabla\times\mathbf{B})_z$, generated through the Amp\`ere--Maxwell relation using the $k \leq 4$ waveform and $E_0 = 10^7$ V m$^{-1}$, showing strongly asymmetric positive and negative excursions within a single optical cycle.   \textbf{(c)} Magnetoelectric-induced bulk magnetic moment $M_x^{\mathrm{ind}}$ as a function of the number of pulse--wait sequences for different waiting times $t_{\mathrm{wait}}$. The induced moment is obtained by averaging the state-resolved magnetoelectric tensor over the full non-equilibrium time-dependent density matrix of the 144-state low-energy manifold generated by the ultrafast dissipative dynamics, and subsequently rescaled to the full irradiation volume containing $4.75\times10^{13}$ molecules under a static readout electric field $E_x = 10^9$ V m$^{-1}$. Dashed curves additionally report the contribution arising solely from the population imbalance $\Delta\rho_{\tau}(t)$ of the maximally toroidal ground-state doublet $|\pm\tau\rangle$, demonstrating that the magnetoelectric response is strongly dominated by the driven polarization of the toroidal ground state. Horizontal dashed lines indicate the estimated $\mu$SQUID sensitivity threshold. For $t_{\mathrm{wait}} = 6.3$ ns and $t_{\mathrm{wait}} = 1$ ns, the threshold is reached after approximately 210 and 217 pulse--wait sequences,  respectively, as indicated by the vertical dashed lines, corresponding to total irradiation times of only $\sim 210$ ns and $\sim 1.37\,\mu$s, respectively. \textbf{(d)} Corresponding accumulated temperature increase $\Delta T$ as a function of the number of pulse--wait sequences for different waiting times $t_{\mathrm{wait}}$. Despite faster thermal accumulation for shorter waiting times, the temperature increase remains modest throughout the experimentally relevant regime, reaching only $\sim 5$ mK and $\sim 9$ mK at the $\mu$SQUID sensitivity threshold for $t_{\mathrm{wait}} = 6.3$ ns and $1$ ns, respectively.}
			\label{fig:6}
	\end{figure}

A realistic implementation of the required temporally asymmetric waveform can be achieved using a superposition of harmonics of a near-infrared laser~\cite{martinez2015boosting}. Specifically, we consider an ultrafast electric field constructed from a fundamental wavelength $\lambda = 1600$ nm and its first three harmonics ($800$ nm, $533$ nm and $400$ nm), yielding the carrier-level waveform
\begin{equation}
	\mathbf{E}(t)=\frac{\hat{\mathbf{z}}}{\pi^2} \frac{\epsilon}{\epsilon - 2} 
	\sum\limits_{k=1}^{4} \frac{(-1)^{k+1}}{k^2}
	\sin\left(\frac{k \pi \left( \epsilon - 2 \right)}{\epsilon}\right)
	\sin\left( \frac{2 \pi k c t}{\lambda} \right),
	\label{eq:sawtoothmaintext}
\end{equation}
already providing an accurate approximation to the target sawtooth asymmetric waveform, as shown in Figure~\ref{fig:6}a,b.

We consider pulse envelopes of $107$ fs, corresponding to $N_{\text{cycles}} = 20$ optical cycles. Using an electric field amplitude $E_0 = 10^7\,\mathrm{V\,m^{-1}}$ over a $20 \ \mu$m spot size (pump fluence $1.42 \ \mu$J cm$^{-2}$), the resulting magnetic-field curl varies asymmetrically within each cycle in the range
\begin{equation}
-0.7 \times 10^{-4} \, \mathrm{T \, nm^{-1}} \leq (\nabla \times \mathbf{B})_z \leq 2.5 \times 10^{-4} \, \mathrm{T \, nm^{-1}}.
\end{equation}

Considering the measured low-temperature heat capacity reported in Figure~\ref{fig:3} ($C/k_B \approx 3$ for $T<1$ K), a laser intensity $I \approx 1.33 \times 10^{11}$ W m$^{-2}$, and a laser spot area $A = \pi \times 10^{-10}$ m$^2$, we estimate the laser-induced heating assuming a weakly absorbing near-transparent regime away from strong optical resonances. A Beer-Lambert estimate using a molar extinction coefficient $\epsilon \sim 1~\mathrm{M^{-1}cm^{-1}}$ yields an absorbed fraction of only $\sim 2.5\%$ of the incident pulse energy (Supplementary Note 7.2). The resulting temperature increase is therefore only $\Delta T_{\mathrm{pulse}}\sim 5\times10^{-2}$ mK per pulse.

Although the thermal diffusion time is estimated to be $t_{\mathrm{diff}}\sim 630$ ns, the present non-resonant weakly absorbing regime differs from conventional resonant pump--probe conditions, where repetition times are typically chosen much longer than $t_{\mathrm{diff}}$ to avoid cumulative heating. Here, however, longer waiting times would also favour relaxation of the accumulated nonequilibrium toroidal polarisation between pulses. The relevant operating regime therefore emerges from a competition between thermal accumulation and toroidal relaxation dynamics. Within the present thermal diffusion model~\cite{carslaw1959conduction,ready1971effects} (Methods~\ref{subsec:tauviadEdt}), pulse--wait sequences with $t_{\mathrm{wait}}=6.3$--$1$ ns ($\sim 160$ MHz--$1$ GHz) already yield a nonequilibrium toroidal population imbalance associated with a detectable $\mu$SQUID magnetoelectric readout signal (Figure~\ref{fig:6}c)~\cite{hasselbach2000microsquid,wernsdorfer2009nanosquid}, while repeated-pulse heating remains bounded despite repetition times about two orders of magnitude shorter than $t_{\mathrm{diff}}$. The $t_{\mathrm{wait}}=6.3$ ns regime provides a conservative operating point with modest thermal accumulation, whereas the $1$ ns limit illustrates a more aggressive accumulation regime where the temperature increase nevertheless remains below the characteristic energy scale of the relevant magnetic excitations. Specifically, the accumulated temperature increase reaches only $\sim 5$ mK and $\sim 9$ mK at the $\mu$SQUID detectability threshold for $t_{\mathrm{wait}}=6.3$ ns and $1$ ns, respectively (Figure~\ref{fig:6}d), more than one order of magnitude below the thermal scale associated with the first low-lying magnetic ($p$-wave/onion-state) excitation at $\sim 0.26$ cm$^{-1}$ ($\sim 370$ mK).

Within this realistic driving scheme, our simulations indicate a nonequilibrium toroidal accumulation per pulse per molecule of approximately $\langle \tau_z \rangle \gtrsim 10^{-8} a_0 \mu_B$ (with $a_0$ and $\mu_B$ the Bohr radius and Bohr magneton), arising from incomplete cancellation of the response within each asymmetric cycle. Explicit numerical propagation of the dissipative many-state dynamics further confirms sustained nonequilibrium accumulation over experimentally relevant pulse--wait sequences and relaxation regimes, with the shortest waiting times approaching an approximately linear growth regime over the timescales considered here (\emph{vide infra}, Figure~\ref{fig:6}c).

To identify an experimental readout mechanism for the nonequilibrium toroidal polarisation, and hence the minimal ultrafast-driving duration required to accumulate a measurable signal, we numerically evaluate the magnetoelectric (ME) response~\cite{spaldin2008toroidal,popov2009anapole,plokhov2011magnetoelectric} associated with the evolving nonequilibrium population distribution. In the presence of toroidal polarisation, the system supports a linear coupling between an external electric field ($E_{\beta}^{\mathrm{ext}}$) and an induced molecular magnetic moment ($M_{\alpha}^{\mathrm{ind}}$),
\begin{equation}
	M_{\alpha}^{\mathrm{ind}} = \sum_\beta \; \alpha_{\alpha\beta}^{\mathrm{ME}} \; E_{\beta}^{\mathrm{ext}},
\end{equation}
where $\alpha_{\alpha\beta}^{\mathrm{ME}}$ is the magnetoelectric tensor.

Here we develop a tailor-made {\em ab initio}-informed computational protocol to evaluate the magnetoelectric response (see Methods~\ref{subsec:magnetoelectric} and Supplementary Note 6), inspired by previous work on toroidal systems~\cite{popov2009anapole, plokhov2011magnetoelectric}.
In this framework, the magnetoelectric tensor associated with each collective Ising configuration of the Fe$_{10}$Dy$_{10}$ wheel included in the ultrafast-driven simulations is computed perturbatively as
\begin{equation}\label{MEtensorMOL}
	\alpha_{\alpha\beta}^{\mathrm{ME}}(\bm{\sigma}) = 
    - \sum_{n} 
    \left[ 
     \frac
     {\langle \bm{\sigma} | P_\alpha | n \rangle \langle n | M_\beta | \bm{\sigma}\rangle}
     {E_n - E_{\bm{\sigma}} }  
     + 
     \frac
     {\langle \bm{\sigma} | M_\beta | n \rangle \langle n | P_\alpha | \bm{\sigma}\rangle}
     {E_n - E_{\bm{\sigma}} } 
     \right],
\end{equation}
where $|\bm{\sigma}\rangle$ denotes one of the 144 low-energy collective Ising states of the ring, with $0\leq E_{\bm{\sigma}}\lesssim 2.5$ cm$^{-1}$ relative to the maximally toroidal ground states $|\pm\tau\rangle$, while the excited states $|n\rangle$ are constructed from local {\em ab initio} crystal-field excitations of the Dy$^{\mathrm{III}}$ ions on the corresponding Ising background. Since the local crystal-field excitation energies ($\gtrsim 100$ cm$^{-1}$) are much larger than the splittings within the 144-state low-energy Ising manifold ($\lesssim 2.5$ cm$^{-1}$), the denominators $(E_n-E_{\bm{\sigma}})$ are well approximated by the local crystal-field excitation energies (Methods~\ref{subsec:magnetoelectric} and Supplementary Note 6). The resulting state-dependent tensors are then averaged over the nonequilibrium populations generated by the dissipative ultrafast dynamics.

The resulting magnetoelectric tensor for the ground toroidal state $|\pm\tau\rangle$ of Fe$_{10}$Dy$_{10}$, expressed in the molecular inertia frame where $z$ is approximately normal to the ring plane while $x$ and $y$ correspond to the long and short in-plane axes of the Dy ring, is
\begin{equation}
	\bm{\alpha}^{\mathrm{ME}}(+\tau)=
	\begin{pmatrix}
		28.185 & -2.10859 & 0.643747 \\
		-12.3051 & 14.1911 & 0.156763 \\
		0.984914 & -0.944577 & 8.99839
	\end{pmatrix}
\end{equation}
(in atomic units, where $1~\mathrm{a.u.~of}~\alpha^{\mathrm{ME}}= 3.8894 \times 10^{-12}\mu_B \,\mathrm{(V/m)^{-1}}$).

Restricting first to the toroidal ground doublet $|\pm\tau\rangle$, which has the largest magnetoelectric response, the most favourable readout channel is obtained by applying the static electric field along the long in-plane axis of the ring and detecting the induced magnetic moment along the same direction, governed by $\alpha_{xx}^{\mathrm{ME}}=28.19$ a.u. To account more accurately for the nonequilibrium population distribution over the full 144-dimensional manifold explored by the ultrafast-driven dynamics, the induced bulk magnetic moment is evaluated as
\begin{equation}
	M_x^{\mathrm{ind}}(t)
	=
	N_{\mathrm{mol}}
	\,\mathrm{Tr}
	\left[
	\bm{\rho}^{\mathrm{NonEq}}(t)
	\,
	\bm{\alpha}_{xx}^{\mathrm{ME}}
	\right]
	E_x^{\mathrm{ext}},
\end{equation}
where $\bm{\rho}^{\mathrm{NonEq}}(t)$ is the nonequilibrium density matrix obtained from the dissipative ultrafast dynamics, $\bm{\alpha}_{xx}^{\mathrm{ME}}$ is the diagonal matrix whose entries are the state-dependent tensor components $\alpha_{xx}^{\mathrm{ME}}(\bm{\sigma})$ associated with the 144 low-energy collective Ising states $\bm{\sigma}$, $N_{\mathrm{mol}}=4.75\times10^{13}$ is the number of molecules within the irradiation volume (Supplementary Note 7.2), and $E_x^{\mathrm{ext}}=10^9$ V\,m$^{-1}$ is the static readout electric field.

Figure~\ref{fig:6}c reports the calculated accumulation of the magnetoelectric readout signal, expressed as the bulk magnetic moment $M_x^{\mathrm{ind}}(t)$ induced by a static electric field $E_x^{\mathrm{ext}}$ acting on the toroidal polarisation generated by the ultrafast pulse train.

For the maximally toroidal ground-state doublet, the expression reduces to
\begin{equation}
	M_x^{\mathrm{GS}}(t)
	=
	N_{\mathrm{mol}}
	\,
	\Delta\rho_{\tau}(t)
	\,
	\alpha_{xx}^{\mathrm{ME}}(+\tau)
	\,
	E_x^{\mathrm{ext}},
\end{equation}
where $\Delta\rho_{\tau}(t)$ is the time-dependent population imbalance between the $+\tau$ and $-\tau$ toroidal ground states accumulated over the pulse train, since the opposite toroidal branches satisfy $\alpha_{xx}^{\mathrm{ME}}(+\tau) = -\alpha_{xx}^{\mathrm{ME}}(-\tau)$.
The corresponding contribution to the accumulated magnetic moment is reported as dashed curves in Figure~\ref{fig:6}c, demonstrating that the nonequilibrium magnetoelectric response remains strongly dominated by the driven population imbalance of the toroidal ground-state manifold $\Delta\rho_{\tau}(t)$, thus providing direct evidence of nonequilibrium toroidal polarization.

From Figure~\ref{fig:6}c it is evident that for both experimentally realistic repetition regimes considered here ($t_{\mathrm{wait}}=6.3$ ns and the more aggressive accumulation-driven $1$ ns limit), the induced bulk magnetic moment exhibits sustained quasi-linear accumulation over at least $300$ pulse--wait sequences, reaching values exceeding the estimated $\mu$SQUID detection threshold of $\sim 10^4\,\mu_B$~\cite{hasselbach2000microsquid, wernsdorfer2009nanosquid}. The threshold is exceeded after approximately $210-217$ pulse--wait sequences, corresponding to total irradiation times of only $\sim1.37\,\mu$s ($t_{\mathrm{wait}}=6.3$ ns) and $\sim210$ ns ($t_{\mathrm{wait}}=1$ ns).
 
Finally, the precise pulse counts and optimal waiting times should be regarded as relaxation-regime dependent, as they are controlled by plausible but phenomenological relaxation rates within a complex dissipative framework. Rather than attempting exhaustive optimisation of the dynamical model, our simulations show that experimentally detectable toroidal accumulation emerges within physically plausible dissipative regimes.

\section{Discussion and Conclusions} \label{sec:3}

	This work shows that Fe$_{10}$Dy$_{10}$ operates in a regime where toroidal polarisation is thermodynamically accessible, dynamically accumulable, and experimentally detectable. Two ingredients are essential: the large molecular toroidal moment, yielding strong coupling to magnetic-field curls, and the ultrafast-driven ratchet mechanism introduced here, which produces an enhancement of the accumulated toroidal polarisation scaling as $|\boldsymbol{\tau}|^4$.
	
	The proposed mechanism is triggered by temporally asymmetric optical fields to generate a net toroidal population imbalance through the interplay of coherent driving and dissipative relaxation, yielding cumulative toroidal polarisation detectable via the magnetoelectric response.
	
	An alternative route would be to generate toroidal polarisation directly via the magnetoelectric effect~\cite{popov2009anapole,plokhov2011magnetoelectric}. In this case, the relevant energy scale is governed by the coupling between electric and magnetic fields mediated by the magnetoelectric tensor. For an isotropic setup, the induced energy splitting is
	\begin{equation}
			\Delta E \approx -\frac{1}{3}\mathrm{Tr}[\boldsymbol{\alpha}^{\mathrm{ME}}]\, \mathbf{E}\cdot\mathbf{B}.
	\end{equation}
		
	For direct comparison with the magnetoelectric splitting estimated by Plokhov, Popov and Zvezdin for a Dy$_3$ single-molecule toroic~\cite{plokhov2011magnetoelectric}, we first consider the same field strengths, namely $E=10^9$ V m$^{-1}$ and $B=1$ T. Using the calculated magnetoelectric tensor of Fe$_{10}$Dy$_{10}$, this gives a splitting of order $	\Delta E \approx 3\times 10^{-2}$ cm$^{-1}$, roughly one order of magnitude larger than the value reported for Dy$_3$~\cite{plokhov2011magnetoelectric}. This comparison should however be regarded as qualitative, since the present estimate is based on a substantially more detailed {\em ab initio}-informed microscopic description of the magnetoelectric tensor than the simplified Dy$_3$ model of Ref.~\cite{plokhov2011magnetoelectric}.
        
     Moreover, magnetic fields of order $1$ T could destabilise the toroidal ground doublet of Fe$_{10}$Dy$_{10}$ by favouring low-lying magnetic states. Restricting to a regime where toroidal character remains preserved ($E=10^9$ V m$^{-1}$ and $B=10$ mT) yields a magnetoelectric splitting of order $10^{-4}$ cm$^{-1}$. This confirms that static magnetoelectric preparation of toroidal polarisation remains possible, but would likely require careful control of field geometry and competing magnetic states.
        
    By contrast, the ultrafast-driven protocol proposed here directly exploits the natural coupling term  $\tau\cdot \nabla\times\mathbf{B}$, avoiding the stabilisation of competing field-induced magnetic states associated with magnetoelectric preparation schemes, while providing a robust route to the accumulation and detection of toroidal polarization in a molecular system.
    More broadly, the present field-curl-driven strategy may provide a useful framework for manipulating toroidal degrees of freedom beyond molecular toroics, including skyrmionic magnetic textures where toroidal and magnetoelectric multipoles have been identified as relevant descriptors of the spin structure~\cite{leonov2015multiply,bhowal2022magnetoelectric,hayami2024multipole}.

\clearpage


\section{Methods} \label{sec:apA}
\subsection{CAHF/CASCI-SO calculations}\label{subsec:cahf}

We performed Configurationally Averaged Hartree Fock (CAHF )/Complete Active Space Configuration Interaction-Spin Orbit (CASCI-SO) calculations on the five symmetry unique [Dy$^{\text{III}}$Fe$^{\text{III}}_2$ (MeTeaH$^{2-}$)(MeTea$^{3-}$)$_2$(CH$_3$O$^-$)$_2$NO$_3^-$]$^{2-}$ (extended) fragments of the Fe$_{10}$Dy$_{10}$ molecular wheel using the in-house developed {\em ab initio} quantum chemistry code CERES~\cite{van2016configuration, calvello2018ceres}, a methodology which allows for the full representation of the spin-orbit coupling problem in the complete basis of the 2002 states arising from full set of $4f^9$ Russell-Saunders terms.   To make the calculations tractable, we replaced the trivalent Fe ions with $3d^{10}$ trivalent Ga ions analogues. We employed the ANO-RCC-VTZP basis set for the Dy atom, ANO-RCC-VDZP for the eight O and N directly coordinating atoms and ANO-RCC-VDZ on all other atoms. A basis set convergence study using higher quality basis sets vindicated our initial finding of energetically well-isolated ground doublets for all Dy$^{\text{III}}$ ions in each fragment. Here below in Table~(\ref{tab:1}) we report the results for the computed $g$-values and the Cartesian components of the principal magnetic axis ($\mathbf{u}$) with largest g-component ($g_Z$) of each symmetry unique Dy$^{\text{III}}$ ion ground doublet. See Supplementary Note 1, Tables S1-S10 for further details about the crystal field spectra, magnetic properties and molecular fragments associated with each Dy(III) ion.

\begin{table}[!ht]
	\caption{\label{tab:1}
		{\em Ab initio} computed $g$-values and the principal magnetic axis ($\mathbf{u}$) associated with the largest g-component ($g_Z$) of each symmetry unique Dy$^{\text{III}}$ ion ground doublet. Lower-case $x,y,z$ denote the molecular Cartesian frame; upper-case $X,Y,Z$ denote the magnetic axes frame.}
	\begin{tabular}{ccccccc}
		\toprule
		\textrm{Fragment}&
		\textrm{$g_{X}$}&
		\textrm{$g_{Y}$}&
		\textrm{$g_{Z}$}&
		\textrm{$(\mathbf{u})_x$}&
		\textrm{$(\mathbf{u})_y$}&
		\textrm{$(\mathbf{u})_z$}\\
		\midrule
		Dy$_{1}$ & 0.05 & 0.11 & 19.55 & -0.7503 & 0.3784 & 0.5422\\
		Dy$_{2}$ & 0.07 & 0.17 & 19.40 & -0.1423 & 0.9042 & -0.4026\\
		Dy$_{3}$ & 0.20 & 0.58 & 19.09 & -0.8837 & -0.0688 & -0.4629\\
		Dy$_{4}$ & 0.22 & 0.68 & 18.85 & 0.1286 & 0.3877 & -0.9128\\
		Dy$_{5}$ & 0.00 & 0.00 & 19.64 & -0.5492 & -0.7106 & -0.4398\\
		\botrule
	\end{tabular}
\end{table}

As it can be seen from Table~\ref{tab:1} and from Tables S1-S5 in the Supplementary Information file, for all Dy$^{\text{III}}$ ions, the results of our calculations revealed energetically well-isolated ground KDs (with first excited KDs residing $60$ cm$^{-1}$ to $150$ cm$^{-1}$ higher in energy) which were comprised of almost pure $\ket{m_J = \pm 15/2}$  total angular momentum states arising from the ${}^{6}H_{15/2}$ ground spin-orbit multiplet. These states displayed rather axial $g$-tensors with average values $\bar{g}_{X} < \bar{g}_{Y} = 0.3$ and $\bar{g}_{Z}= 19.3$ (see Table~\ref{tab:1} in the Methods section).  The {\em ab initio} principal magnetic Z-axes $\mathbf{u}_i$ for each Dy$^{\text{III}}$ are given in Table \ref{tab:1} and are shown as orange arrows centred on each Dy$^{\text{III}}$ ion in Figure \ref{fig:1}.

\subsection{Broken symmetry density functional theory calculations}\label{subsec:DFT}

To determine the sign and magnitude of the fifteen symmetry-unique intramolecular exchange coupling constants between nearest- and next-nearest-neighbour paramagnetic ions, we employed broken-symmetry density functional theory (DFT) calculations~\cite{noodleman1981valence} on isostructural analogues of the Fe$^{\text{III}}_i$-Dy$^{\text{III}}_i$-Fe$^{\text{III}}_{i+1}$ fragments of the molecular wheel with appropriate metal substitutions {\em vide infra}. Calculations were performed with ORCA 5.0~\cite{neese2017software} using the TPSSh functional~\cite{tao2003climbing} and SARC-DKH-TZVP basis sets~\cite{pantazis2009all} for all atoms.

For the $J_A$ coupling constants, we converged high-spin and broken-symmetry calculations for each of the five symmetry-unique Ga$^{\text{III}}_{i-1}$-Gd$^{\text{III}}_i$-Fe$^{\text{III}}_i$ fragments, where Fe$^{\text{III}}_{i-1}$ was replaced by diamagnetic Ga$^{\text{III}}$ and Dy$^{\text{III}}_i$ by orbitally non-degenerate Gd$^{\text{III}}_i$. The resulting couplings were scaled using $S^{\text{Dy}}/S^{\text{Gd}} = 5/7$ and extracted via Yamaguchi's formula~\cite{yamaguchi1986molecular}. The same strategy was employed for $J_B$ and $J_C$ using the fragments Fe$^{\text{III}}_{i-1}$-Gd$^{\text{III}}_i$-Ga$^{\text{III}}_i$ and Fe$^{\text{III}}_{i-1}$-La$^{\text{III}}_i$-Fe$^{\text{III}}_i$, respectively.

\subsection{The interaction Hamiltonian}\label{subsec:ham}

Within a given Ising configuration $\boldsymbol{\sigma}$, the exchange Hamiltonian is
\begin{equation} \label{eq:3}
	H_{\text{ex}}(\boldsymbol{\sigma}) = - \sum\limits_{i=1}^{10} \mathbf{S}_{i}^{\text{Fe}} \cdot \left( J^{i}_A \mathbf{S}_{i}^{\text{Dy}} + J^{i}_B \mathbf{S}_{i+1}^{\text{Dy}} + J^{i,i+1}_C \mathbf{S}_{i+1}^{\text{Fe}} \right),
\end{equation}
where $J_A$ and $J_B$ are nearest-neighbour Fe$^{\text{III}}$-Dy$^{\text{III}}$ exchange couplings and $J_C$ is the next-nearest-neighbour Fe$^{\text{III}}$-Fe$^{\text{III}}$ coupling.

The Dy$^{\text{III}}$ Ising spins are $\mathbf{S}_{i}^{\text{Dy}} = \frac{5}{2} \sigma_i \mathbf{u}_i$, thus fixed by the Ising configuration $\boldsymbol{\sigma}$ (where $\mathbf{u}_i$ the ab initio magnetic axis of the $i^{th}$ ion as reported in Table~\ref{tab:1}), whereas $\mathbf{S}^{\text{Fe}}_{i}$ acts on the local $S^{\text{Fe}}=5/2$ true quantum spin manifold of each Fe$^{\text{III}}$ ion.

Intramolecular dipolar interactions are described by
\begin{equation} \label{eq:4}
	H_{\text{dip}}(\boldsymbol{\sigma}) = \frac{\mu_0}{4 \pi} \sum\limits_{ij} \frac{\mathbf{M}_{i} \cdot \mathbf{M}_{j}}{\abs{\mathbf{R}_{ij}}^3} - 3 \frac{\left( \mathbf{M}_{i} \cdot \mathbf{R}_{ij} \right) \left( \mathbf{M}_{j} \cdot \mathbf{R}_{ij} \right)}{\abs{\mathbf{R}_{ij}}^5}
\end{equation}
where $\mathbf{R}_{ij}=\mathbf{r}_i-\mathbf{r}_j$ and $\mathbf{M}_{i}^{\text{Fe}} = g \mu_B \mathbf{S}_{i}^{\text{Fe}}$ with $g=2$, while $\mathbf{M}_i^{\text{Dy}} = \frac{1}{2}\mu_B g_{zz}^i \sigma_i \mathbf{u}_i$.
Only dipolar interactions within each Dy-Fe-Dy fragment are retained. Magnetic dipole-dipole coupling between Fe$^{\text{III}}$ next-nearest neighbours is approximately $\mu_0 (\mu_B g S^{\text{Fe}})^2 /4 \pi R^3 \sim 0.05$ cm$^{-1}$ and thus can be safely neglected as can dipolar couplings to further removed Fe$^{\text{III}}$ ions in the ring.

The Zeeman and toroidal interactions are
\begin{equation} \label{eq:5}
	\begin{aligned}
		H_{\text{Zee}}(\boldsymbol{\sigma}) &= \sum\limits_{i=1}^{10} \left( \mathbf{M}_{i}^{\text{Fe}} + \mathbf{M}_{i}^{\text{Dy}} \right) \cdot \mathbf{B}\\[4mm]
		&= \mathbf{M} \cdot \mathbf{B}
	\end{aligned}
\end{equation}
and 
\begin{equation} \label{eq:6}
	\begin{aligned}
		H_{\text{Tor}}(\boldsymbol{\sigma}) &= \sum\limits_{i=1}^{10} \left( \mathbf{r}^{\text{Fe}}_{i} \times \mathbf{M}^{\text{Fe}}_{i} + \mathbf{r}^{\text{Dy}}_{i} \times \mathbf{M}^{\text{Dy}}_{i} \right) \cdot \left( \nabla \times \mathbf{B} \right)\\[4mm]
		&= \boldsymbol{\tau} \cdot \left( \nabla \times \mathbf{B} \right).
	\end{aligned}
\end{equation}
The toroidal coupling provides direct access to thermodynamic toroidal response properties via free-energy derivatives, as discussed in Section~\ref{sec:MaxTorState}.


\subsection{Perturbative treatment of Fe-Fe exchange interaction}\label{subsec:fefept}

Since our broken symmetry DFT calculations indicated systematically a smaller value of $J_C$ with respect to Fe$^{\text{III}}$-Dy$^{\text{III}}$ exchange and magnetic dipole coupling (see Table \ref{tab:2}), we further treat the antiferromagnetic Fe$^{\text{III}}$-Fe$^{\text{III}}$ exchange as a first order correction to the zeroth order energies $E^{(0)}_{\boldsymbol{\sigma} \boldsymbol{\lambda}(\boldsymbol{\sigma})}$ obtained from the following zeroth order block-diagonal Hamitonian ($H_0(\boldsymbol{\sigma})$), in which the problem is partitioned into a sum of local Fe$^{\text{III}}$ Hamiltonians as:
\begin{equation} \label{eq:7}
	\begin{aligned}
		H_0(\boldsymbol{\sigma}) &= H(\boldsymbol{\sigma}) - \left( - \sum\limits_{i=1}^{10} J_C^{i,i+1} \mathbf{S}^{\text{Fe}}_i \cdot \mathbf{S}^{\text{Fe}}_{i+1} \right)\\
		&= \sum\limits_{i=1}^{10} h_i(\sigma_i, \sigma_{i+1})
	\end{aligned}
\end{equation}
where each $h_i(\sigma_i, \sigma_{i+1})$ details the exchange, magnetic dipole and external field interactions of the Fe$^{\text{III}}$ ion at site $i$. Since $\left[ h_i(\sigma_i, \sigma_{i+1}), h_j(\sigma_j, \sigma_{j+1}) \right] = 0$, each local Fe$^{\text{III}}$ Hamiltonian can be diagonalised on the local six dimensional $\ket{m_i}$ basis of each Fe$^{\text{III}}$ site to give
\begin{equation} \label{eq:8}
	h_i(\sigma_i, \sigma_{i+1}) \ket{\lambda_i^{\sigma_i, \sigma_{i+1}}} = \epsilon_{\lambda_i}(\sigma_i, \sigma_{i+1}) \ket{\lambda_i^{\sigma_i, \sigma_{i+1}}}
\end{equation}
where the $\lambda_i$'s index the six energy eigenstates at each site. Naturally, the ring eigenstates for a given dysprosium configuration $\ket{\boldsymbol{\sigma}}$ and Fe$^{\text{III}}$ excitation pattern $\boldsymbol{\lambda}(\boldsymbol{\sigma})$ are product states of the local Fe$^{\text{III}}$ eigenstates at each site $\ket{\boldsymbol{\lambda}(\boldsymbol{\sigma})} = \ket{\lambda_1^{\sigma_1, \sigma_{2}}} \otimes \dots \otimes \ket{\lambda_{10}^{\sigma_{10}, \sigma_{1}}}$ with total energy 
\begin{equation} \label{eq:9}
	E^{(0)}_{\boldsymbol{\sigma}, \boldsymbol{\lambda}(\boldsymbol{\sigma})} = \sum\limits_{i=1}^{10} \epsilon_{\lambda_i}(\sigma_i, \sigma_{i+1}).
\end{equation}
hence the perturbative correction due to antiferromagnetic Fe-Fe exchange coupling is evaluated as:
\begin{equation} \label{eq:10}
	E_{\boldsymbol{\sigma}, \boldsymbol{\lambda}(\boldsymbol{\sigma})}^{(1)} = - \sum\limits_{i=1}^{10} J_C^{i,i+1} \langle \mathbf{S}^{\text{Fe}}_i \rangle_{\lambda_i} \cdot \langle \mathbf{S}^{\text{Fe}}_{i+1} \rangle_{\lambda_{i+1}}
\end{equation}
where the expectation values are simply the diagonal matrix elements $\langle \mathbf{S}^{\text{Fe}}_i \rangle_{\lambda_i} = \bra{\lambda_i^{\sigma_i, \sigma_{i+1}}} \mathbf{S}^{\text{Fe}}_{i} \ket{\lambda_i^{\sigma_i, \sigma_{i+1}}}$.

The free energy and response properties for the whole system are finally computed making use of the transfer matrix approach as detailed in the next subsection~\ref{subsec:transfermat}.
The effect of spin frustration obtained by this perturbative approach on the ring's spectrum and states, is discussed in the Methods' subsection~\ref{subsec:spectrum}, and also in the main text.

\subsection{Free energy calculation}\label{subsec:transfermat}

To compute thermodynamic quantities (e.g. magnetisation, magnetic susceptibility) from our model requires knowledge of the partition function $Z$ which, as has already been noted, is the sum of $2^{10} \times 6^{10}$ terms. Rather than carrying out this sum explicitly, we show illustrate how the Fe$_{10}$Dy$_{10}$ partition function can be rewritten as the trace over a product of ten $24 \times 24$ dimensional transfer matrices~\cite{van2010dysprosium} allowing for the efficient calculation of the ring free energy and, subsequently, thermodynamic quantities of interest.

The partition function for Fe$_{10}$Dy$_{10}$ is the sum of the exponentiated energies defined in Eq. (\ref{eq:9}) and Eq. (\ref{eq:10}) for each possible Ising configuration of the Dy$^{\text{III}}$ magnetic axes $\boldsymbol{\sigma}$ and Fe$^{\text{III}}$ excitation pattern $\boldsymbol{\lambda}(\boldsymbol{\sigma})$
\begin{equation} \label{eq:11}
	\begin{aligned}
		Z &= \sum\limits_{\boldsymbol{\sigma}, \boldsymbol{\lambda}(\boldsymbol{\sigma})}^{6^{10} \times 2^{10}} e^{- \beta \left(E_{\boldsymbol{\sigma}, \boldsymbol{\lambda}(\boldsymbol{\sigma})}^{(0)} + E_{\boldsymbol{\sigma}, \boldsymbol{\lambda}(\boldsymbol{\sigma})}^{(1)} \right)}\\[4mm]
		&= \sum\limits_{\boldsymbol{\sigma}, \boldsymbol{\lambda}(\boldsymbol{\sigma})}^{6^{10} \times 2^{10}} \prod\limits_{i=1}^{10} e^{- \beta \Theta(\lambda_i \sigma_i \sigma_{i+1}; \lambda_{i+1} \sigma_{i+1} \sigma_{i+2})}
	\end{aligned}
\end{equation}
where in the second line we have collected energies corresponding to each Fe$_i^{\text{III}}$-Fe$_{i+1}^{\text{III}}$ `bond' in the term
\begin{equation} \label{eq:12}
	\begin{aligned}
		& \Theta(\lambda_i \sigma_i \sigma_{i+1}; \lambda_{i+1} \sigma_{i+1} \sigma_{i+2}) = \frac{1}{2} \left[ \epsilon_{\lambda_i}(\sigma_i, \sigma_{i+1}) \ + \right.\\[4mm]
		& \qquad \left. \epsilon_{\lambda_{i+1}}(\sigma_{i+1}, \sigma_{i+2}) \right] - J_C^{i,i+1} \langle \mathbf{S}^{\text{Fe}}_{i} \rangle_{\lambda_i} \cdot \langle \mathbf{S}^{\text{Fe}}_{i+1} \rangle_{\lambda_{i+1}}.
	\end{aligned}
\end{equation}
Defining the matrices $A^{i,i+1}$ with matrix elements
\begin{equation} \label{eq:13}
	\begin{aligned}
		A^{i,i+1}_{\lambda_i \sigma_i \sigma_{i+1}; \lambda_{i+1} \sigma_{i+1} \sigma_{i+2}} &= \delta_{\sigma_{i+1} \sigma'_{i+1}}\\[4mm]
		& \times e^{-\beta \Theta(\lambda_i \sigma_i \sigma_{i+1}; \lambda_{i+1} \sigma'_{i+1} \sigma_{i+2})}
	\end{aligned}
\end{equation}
where $\delta_{ij}$ is the Kronecker delta, we identify the sum of products on the last line of Eq. (\ref{eq:11}) with the trace of a product of matrices $A^{1,2} \dots A^{10,1}$. Clearly from the definition of the compound indices $\lambda_i \sigma_i \sigma_{i+1}$, the $A^{i,i+1}$ matrices are just $(6 \times 2 \times 2)^2 = 24 \times 24$ dimensional.

Thus the Fe$_{10}$Dy$_{10}$ free energy (F) can be expressed as
\begin{equation} \label{eq:14}
	\begin{aligned}
		F &= \frac{-1}{\beta} \log \left( Z \right)\\
		&= \frac{-1}{\beta} \log \left( \Tr \left[ A^{1,2} \dots A^{10,1} \right] \right).
	\end{aligned}
\end{equation}
which is certainly computationally more palatable than a sum over $2^{10} \times 6^{10} \simeq 62$ billion states. To calculate thermodynamic quantities of interest we take the appropriate derivatives of Eq. (\ref{eq:14}) as discussed in Section~\ref{sec:ComparisonExperiments}.

\subsection{Non-equilibrium Toroidal Dynamics under Pulsed Electric Driving}\label{subsec:tauviadEdt}

To rationalise the dynamical accumulation of toroidal polarisation, we first consider an effective two-level model formed by two time- and parity-reversed toroidal states carrying opposite toroidal moments $\pm\boldsymbol{\tau}$. The system is driven by the magnetic-field curl generated by a temporally asymmetric electric waveform,
\begin{equation}
	\nabla \times \mathbf{B}(t)
	=
	\frac{1}{c^2}\frac{\partial \mathbf{E}(t)}{\partial t},
\end{equation}
which couples to the two-level system through
\begin{equation}
	\mathcal{H}(t)
	=
	\boldsymbol{\tau}\cdot(\nabla\times\mathbf{B}(t))\,\sigma_z .
\end{equation}
Dissipative population transfer between the two toroidal branches is described by a Lindblad equation,
\begin{equation}
	\dot{\rho}
	=
	-\frac{i}{\hbar}[\mathcal{H}(t),\rho]
	+
	\kappa g(t)\mathcal{L}[\sigma_+]
	+
	\kappa h(t)\mathcal{L}[\sigma_-],
\end{equation}
where $\mathcal{L}[O]=O\rho O^\dagger-\frac12O^\dagger O\rho-\frac12\rho O^\dagger O$, and the rates $g(t)$ and $h(t)$ obey detailed balance with respect to the instantaneous toroidal splitting
\begin{equation}
	\Delta(t)=2\boldsymbol{\tau}\cdot(\nabla\times\mathbf{B}(t)).
\end{equation}
Expanding the density matrix perturbatively in the dissipative rate, $\rho(t)=\sum_r\kappa^r\rho^{(r)}(t)$, shows that coherent driving alone only generates phases, while the combination of dissipation and temporal asymmetry produces a population imbalance. For the asymmetric sawtooth waveform defined in Supplementary Note 7, the leading contribution after one optical period $T_l$ is
\begin{equation}
	\langle \tau (t) \rangle
	=
	\frac{16 \kappa \abs{\boldsymbol{\tau}}^4}{T_l^3} \left( \frac{E_0}{c^2} \right)^3 \frac{\epsilon^3(\epsilon - 4)}{(\epsilon - 2)^2} t + \mathcal{O}(\kappa^2),
\end{equation}
where $E_0$ is the electric-field amplitude and $\epsilon$ controls the temporal asymmetry. This expression shows that accumulation vanishes for the symmetric waveform, $\epsilon=4$, and scales as $|\boldsymbol{\tau}|^4$, motivating large molecular rings with large intrinsic toroidal moments.

For Fe$_{10}$Dy$_{10}$, we then replace the two-level model by a reduced low-energy Ising basis. The full Hilbert space is made tractable by freezing the Fe$^{\mathrm{III}}$ spin states in their zero-curl ground configurations and retaining only Dy$^{\mathrm{III}}$ Ising configurations with energies $\leq 2.5$ cm$^{-1}$, appropriate for low-temperature and weak-driving dynamics. The populations $\rho_m$ of the retained configurations obey the time-dependent master equation
\begin{equation}
	\frac{d\rho_m}{dt}
	=
	\sum_{k\neq m}
	\left[
	W^{k\to m}(t)
	+
	\Omega^{k\leftrightarrow m}(t)
	\right]\rho_k
	-
	\rho_m
	\sum_{k\neq m}
	\left[
	W^{m\to k}(t)
	+
	\Omega^{k\leftrightarrow m}(t)
	\right],
\end{equation}
with phonon-assisted transitions
\begin{equation}
	W^{m\to k}(t)
	=
	\left(
	\Gamma_1\delta_{mk}^{1-\mathrm{Dy}}
	+
	\Gamma_2\delta_{mk}^{2-\mathrm{Dy}}
	+
	\Gamma_3\delta_{mk}^{3-\mathrm{Dy}}
	\right)
	\frac{\left[E_k(t)-E_m(t)\right]^3}
	{\exp\{\beta[E_k(t)-E_m(t)]\}-1},
\end{equation}
and tunnelling rates
\begin{equation}
	\Omega^{m\leftrightarrow k}(t)
	=
	\left(
	\gamma_1\delta_{mk}^{1-\mathrm{Dy}}
	+
	\gamma_2\delta_{mk}^{2-\mathrm{Dy}}
	+
	\gamma_3\delta_{mk}^{3-\mathrm{Dy}}
	\right)
	\frac{\xi}
	{\left([E_k(t)-E_m(t)]/\hbar\right)^2+\xi^2}.
\end{equation}
Here $\delta_{mk}^{r-\mathrm{Dy}}=1$ if configurations $m$ and $k$ differ by $r$ Dy moment flips and zero otherwise, while $E_k(t)$ is the instantaneous energy of configuration $k$ in the applied curling field. For our simulations we adopt the leading-order spin-phonon coupling constant $\Gamma_1 = 10^2$ cm$^3$ s$^{-1}$~\cite{briganti2021complete,zhu2025chiral}, and the quantum tunnelling parameters $\gamma_1 = 10^{16}$ s$^{-2}$ and $\xi = 10^{10}$ s$^{-1}$, corresponding to an on-resonance tunnelling rate $\gamma_1/\xi \sim 10^6$ s$^{-1}$~\cite{ding2018field}, compatible with tunnelling rates reported for other dysprosium single-molecule magnets~\cite{vignesh2017ferrotoroidic, ashtree2021tuning}. We further imposed a reasonable hierarchy on the higher-order multi-flip spin-phonon rate constants, $\Gamma_2 = 10^{-2}\Gamma_1$ and $\Gamma_3 = 10^{-4}\Gamma_1$. A similar hierarchy was adopted for the multi-site quantum tunnelling rate constants, $\gamma_2 = 10^{-1}\gamma_1$ and $\gamma_3 = 10^{-2}\gamma_1$, which conservatively represent the fastest identified two-site and three-site tunnelling processes in Fe$_{10}$Dy$_{10}$. The relative efficiency of such processes was estimated from products of the transverse {\em ab initio} g-tensor components reported in Supplementary Tables S1--S5, whose strong variability reflects the broad distribution of local transverse anisotropies and tunnelling efficiencies in this system.


The driving field is generated from a truncated Fourier representation of a temporally asymmetric sawtooth waveform,

\begin{equation}
	\mathbf{E}(t)
	=
	\frac{\hat{\mathbf{z}}}{\pi^2}
	\frac{\epsilon}{\epsilon-2}
	\sum_{k=1}^{4}
	\frac{(-1)^{k+1}}{k^2}
	\sin\!\left[
	\frac{k\pi(\epsilon-2)}{\epsilon}
	\right]
	\sin\!\left(
	\frac{2\pi kct}{\lambda}
	\right),
\end{equation}
implemented using the fundamental wavelength $\lambda=1600$ nm and its first harmonics, as described in Supplementary Note 7. The toroidal polarisation at time $t$ is then evaluated as
\begin{equation}
\langle \tau(t)\rangle=\sum_m \rho_m(t)\tau_m .
\end{equation}

The experimentally detectable signal is obtained by combining the nonequilibrium populations generated by the ultrafast dynamics with the state-dependent magnetoelectric tensors derived above. The nonequilibrium populations $\rho_m(t)$ are assembled into the diagonal density matrix $\bm{\rho}^{\mathrm{NonEq}}(t)$, while the corresponding state-dependent magnetoelectric tensor components $\alpha_{xx}^{\mathrm{ME}}(\bm{\sigma})$ for the selected 144 Ising states $\bm{\sigma}$ define the diagonal matrix $\bm{\alpha}_{xx}^{\mathrm{ME}}$. The induced bulk magnetic moment is then evaluated as

\begin{equation}
	M_x^{\mathrm{ind}}(t)=
	N_{\mathrm{mol}}
	\,\mathrm{Tr}\left[
    \bm{\rho}^{\mathrm{NonEq}}(t)\,\bm{\alpha}_{xx}^{\mathrm{ME}}
	\right] E_x^{\mathrm{ext}},
\end{equation}
where $E_x^{\mathrm{ext}}$ is the static readout electric field applied along the long in-plane axis of the ring.

For a laser volume containing
\begin{equation}
	N_{\mathrm{mol}}
	=
	\frac{\rho_{\mathrm{Fe}_{10}\mathrm{Dy}_{10}}N_A V_{\mathrm{laser}}}
	{M_{\mathrm{Fe}_{10}\mathrm{Dy}_{10}}}
	\approx
	4.75 \times 10^{13},
\end{equation}
the induced magnetic moment becomes experimentally detectable within the accumulation regime discussed in the main text.

To evaluate the thermal load induced by the non-resonant laser pulse train, we adopt the standard linear superposition framework for multi-pulse heat accumulation~\cite{ready1971effects,eaton2005heat}. For an initial Gaussian temperature profile of characteristic radius $R$, the solution of the three-dimensional heat diffusion equation gives a peak temperature excess at the centre of the laser spot ($r=0$) that decays as~\cite{carslaw1959conduction}
\begin{equation}
\Delta T(t)=
\Delta T_0
\left(
1+\frac{t}{t_{\mathrm{diff}}}
\right)^{-3/2},
\end{equation}
where $\Delta T_0$ is the instantaneous temperature increase induced by a single pulse and $t_{\mathrm{diff}}$ is the characteristic thermal diffusion time.

For a sequence of $N_{\mathrm{pulses}}$ pulses separated by a waiting time $t_{\mathrm{wait}}$, the accumulated temperature increase is obtained by linear superposition of the residual contributions from all previous pulses,
\begin{equation}
\Delta T(N_{\mathrm{pulses}})
=
\Delta T_0
\sum_{k=0}^{N_{\mathrm{pulses}}-1}
\left(
1+k\frac{t_{\mathrm{wait}}}{t_{\mathrm{diff}}}
\right)^{-3/2}.
\end{equation}

In the experimentally relevant regime considered here ($t_{\mathrm{wait}}\ll t_{\mathrm{diff}}$), the sum is well approximated by the continuous-limit expression
\begin{equation}
\Delta T(N_{\mathrm{pulses}})
\simeq
\Delta T_0
\frac{2t_{\mathrm{diff}}}{t_{\mathrm{wait}}}
\left[
1-
\left(
1+
N_{\mathrm{pulses}}
\frac{t_{\mathrm{wait}}}{t_{\mathrm{diff}}}
\right)^{-1/2}
\right].
\end{equation}

The parameters $\Delta T_0$ and $t_{\mathrm{diff}}$ are obtained from the absorbed pulse energy, measured low-temperature heat capacity and estimated thermal diffusivity, as detailed in Supplementary Note 7. Full details of the perturbative two-level derivation, rate parametrisation, pulse shaping, numerical integration, magnetoelectric readout and heating estimates are given in Supplementary Note 7.




\subsection{Magnetoelectric Coupling Tensor}\label{subsec:magnetoelectric}

Here we derive the magnetoelectric (ME) response of a polynuclear rare-earth complex within perturbation theory as the leading mixed derivative of the energy with respect to electric and magnetic fields. In toroidal systems this response is linear and corresponds to the lowest-order spin-electric coupling~\cite{spaldin2008toroidal,popov2009anapole}. Owing to the weakness of 4f--4f exchange, the response is dominated by single-ion contributions~\cite{plokhov2011magnetoelectric}.

Within the single-ion mechanism, the electric field couples the local 4f states of each Ln(III) ion to the 5d manifold of the same ion through the odd-harmonic components of the effective crystal field, in close analogy with Judd–Ofelt theory for optical transitions in lanthanide systems~\cite{wybourne2007opticalspectroscopy}.

Building on Ref.~\cite{popov2009anapole}, we retain the same second-order perturbative structure and constant $4f$--$5d$ energy-denominator approximation, but restrict the $4f$ states to the Hund's-rule ground Russell--Saunders term and its spin-orbit multiplet. Within this manifold, crystal-field matrix elements are evaluated directly via the Wigner--Eckart theorem, eliminating the need for coefficients of fractional parentage while remaining exact for the quantities of interest.



Moreover, we generalise previous symmetry-restricted models~\cite{popov2009anapole,plokhov2011magnetoelectric} by explicitly constructing the internal electrostatic potential from effective point charges placed on the experimental Fe$_{10}$Dy$_{10}$ structure, in the spirit of the Magellan electrostatics approach~\cite{chilton2013electrostatic}. 

We further retain the full hierarchy of odd-harmonic crystal-field contributions ($k=1,3,5$), corresponding to the coupling of the electric field and its spatial derivatives to the dipole ($k=1$), octupole ($k=3$) and dotriacontapole ($k=5$) moments of the 4f–5d overlap distribution, thereby extending previous symmetry- and dipole-restricted ($k=1$) treatments~\cite{popov2009anapole,plokhov2011magnetoelectric}.

Finally, the resulting local electric and magnetic Hamiltonians are projected onto the {\em ab initio} decomposition of a representative Dy(III) local wavefunction (Supplementary Note 6), assumed transferable across the ten Dy sites consistently with the approximate level of theory adopted. The resulting local response is then projected onto the collective toroidal states of the molecule in the presence of an external magnetic field.

\paragraph*{Effective electric Hamiltonian within the $4f$ manifold.}
We start from the electric perturbations acting on the $4f^N$ shell of a single lanthanide ion~\cite{wybourne2007opticalspectroscopy, popov2009anapole},
\begin{equation}
	\hat V = \hat V_1 + \hat V_2,
\end{equation}
where
\begin{equation}
\label{V1V2secondQ}
\hat{V}_i = \sum_{b,c} v^{(i)}_{bc} a^{\dagger}_{b}a_{c}, \;\; i=1,2
\end{equation}
where $a^{\dagger}_{b}$ ($a_{c}$) are the creation (annihilation) operators varying the occupation number of the associated one-electron spin-orbital basis $\{\phi_b\}$, while the one-electron matrix elements $v^{(1)}_{bc}$ read:
\begin{equation}
\label{v1secQ}
v^{(1)}_{bc} = \sum_{q=-1}^{+1} (-1)^q E_{-q}^{\mathrm{ext}} \left\langle \phi_b \left|  r C^{(1)}_q(\bf{r}) \right| \phi_c \right\rangle ,
\end{equation}
and, likewise, the matrix elements $v^{(2)}_{bc}$ read:
\begin{equation}
\label{v2secQ}
v^{(2)}_{bc} = \sum_k^{1,3,5}\sum_{q=-k}^{+k} A^{\mathrm{CF}}_{kq} \left\langle \phi_b \left|  r^k C^{(k)}_q(\bf{r}) \right| \phi_c \right\rangle,
\end{equation}
where $E_{-q}^{\mathrm{ext}}$ are the spherical components of the external electric field, $C^{(k)}_q(\mathbf{r})$ are Wybourne irreducible tensor operators built from an open-shell electron position $\mathbf{r}$, and $A^{\mathrm{CF}}_{kq} = e \sum_n (Q_n/R_n^{k+1})C^{(k)*}_q(\mathbf{R}_n)$ are the irreducible tensor components of the crystal field potential generated by point charges $eQ_n$ placed at position $\mathbf{R}_n$.

Both operators $\hat{V}_i$ have vanishing matrix elements within the $4f^N$ space, and contribute only via virtual excitations to states from opposite-parity configurations.

Using quasi-degenerate perturbation theory and restricting the opposite-parity intermediate space to $4f^{N-1}5d^1$ (see Supplementary Note 6 for details), after explicit elimination of the intermediate space, we arrive at the following effective Hamiltonian:
\begin{equation}
\label{heffFIRST}
\hat{H}_{\mathrm{eff}} = -\frac{1}{W_{fd}} \hat{P}_0 \left( \hat{V}_1^{-} \hat{V}_2^{+} + \hat{V}_2^{-}\hat{V}_1^{+} \right)\hat{P}_0,
\end{equation}
where $\hat{P}_0$ is a projector onto the $4f^{N}$ space (see Supplementary Note 6), $W_{fd}$ is an average $4f$–$5d$ excitation energy~\cite{wybourne2007opticalspectroscopy,popov2009anapole,plokhov2011magnetoelectric}, and:
\begin{eqnarray*}
\hat{V}_i^{(-)} &=& \sum_\sigma^{\alpha,\beta}\sum_{m_f\in 4f}\sum_{m_d\in 5d} v^{(i)}_{m_f m_d}\,a^\dagger_{m_f,\sigma}a_{m_d,\sigma} \\
\hat{V}_i^{(+)} &=& \sum_\sigma^{\alpha,\beta}\sum_{m_f\in 4f}\sum_{m_d\in 5d} v^{(i)}_{m_d m_f}\,a^\dagger_{m_d,\sigma}a_{m_f,\sigma}.
\end{eqnarray*}
Further manipulation of Eq.~(\ref{heffFIRST}) yields an explicit one-electron effective Hamiltonian acting solely within the $4f$ shell,
\begin{equation}
\label{heffSECOND}
	\hat H_{\mathrm{eff}}^{(\mathrm{el})}
	=
	\sum_{m_1,m_2\in 4f}
	G_{m_1 m_2}\,
	\hat E_{m_1 m_2},
\end{equation}
with $\hat E_{m_1 m_2} = a^\dagger_{m_1,\alpha}a_{m_2,\alpha}+a^\dagger_{m_1,\beta}a_{m_2,\beta}$ the second-quantization singlet excitation operators, and with coupling matrix elements:
\begin{equation}
\label{Gm1m2}
G_{m_1m_2}
=
\sum_{\mu=-1}^{+1}(-1)^\mu E^{\mathrm{ext}}_{-\mu}
\sum_{g=1,3,5}\sum_{\gamma=-g}^{+g}
G_{fd}^{(g)} A^{\mathrm{CF}}_{g\gamma} \,
\left(
O_{\mu\gamma}^{1,g}(m_1,m_2)
+
O_{\gamma\mu}^{g,1}(m_1,m_2)
\right),
\end{equation}
where
\begin{equation}
\label{RadialCoupling}
G_{fd}^{(g)}=-e^2\frac{\langle r\rangle_{fd}\langle r^g\rangle_{fd}}{W_{fd}}
\end{equation}
is the effective coupling constant proportional to the product of the radial transition dipole $\langle r\rangle_{fd}$ and the radial transition multipoles $\langle r^g\rangle_{fd}$ of the crystal field harmonics of rank $g=1,3,5$, and
\begin{equation}
O_{\mu\gamma}^{1,g}(m_1,m_2)
=
\sum_{m_d=-2}^{+2}
\mel{3m_1}{C_\mu^{(1)}}{2m_d}
\mel{2m_d}{C_\gamma^{(g)}}{3m_2},
\end{equation}
with an analogous definition for $O_{\gamma\mu}^{g,1}(m_1,m_2)$.
The three radial coupling constants Eq.~(\ref{RadialCoupling}) were approximated here with the radial integrals of second-order Judd--Ofelt for Dy(III) tabulated by Wybourne and Smentek in Table 21-9 at page 294 of~\cite{wybourne2007opticalspectroscopy}, which already account for the excitation energy gap. The three values used here, in atomic units (a.u.), are thus $G^{(1)}_{fd} = -0.2949$ a.u., $G^{(3)}_{fd} = -1.0005$ a.u., and $G^{(5)}_{fd}=-4.8183$ a.u. 

Using standard angular momentum recoupling arguments (see Supplementary Note 6), we further find that
\begin{equation}
\label{Om1m2}
O_{\mu\gamma}^{1,g}(m_1,m_2)
=
\sum_{K=2,4,6}\Omega_g^{(K)}
\sum_{Q=-K}^{+K}
\cg{1}{\mu}{g}{\gamma}{K}{Q}
(-1)^{3-m_1}
\threej{3}{K}{3}{-m_1}{Q}{m_2},
\end{equation}
where 
\begin{equation}
\Omega^{(k)}_{g} = (-1)^{k + g - 3} \, \sqrt{2k + 1} \,
\begin{Bmatrix}
3 & 3 & k \\
1 & g & 2
\end{Bmatrix}
\, \langle 3 \| C^{(1)} \| 2 \rangle \,
\langle 2 \| C^{(g)} \| 3 \rangle ,
\end{equation}
where $\langle 2 \| C^{(g)} \| 3 \rangle$ are standard reduced matrix elements of spherical tensor operators between $f$ and $d$ single-electron orbital angular momentum eigenstates~\cite{varshalovich1988quantum}.

Substituting Eq.~(\ref{Om1m2}) and Eq.~(\ref{Gm1m2}) into Eq.~(\ref{heffSECOND}), and introducing the coupled tensor operators 
\[
\hat U_Q^{(K)} = \sum_{m_1,m_2}(-1)^{3-m_1}
\threej{3}{K}{3}{-m_1}{Q}{m_2}\hat E_{m_1m_2},
\]
we obtain (see Supplementary Note 6) the effective Hamiltonian of the $I^{th}$ Dy(III) ion:
\begin{equation}
\label{HeffTHIRD}
\hat H_{\mathrm{eff}}^{(\mathrm{el})}(I)
=
2 \sum_{K=2,4,6}\sum_{Q=-K}^{+K}
B_{KQ}^{(12)}(I)\; \hat U_Q^{(K)},
\end{equation}
where
\begin{equation}
B_{KQ}^{(12)}(I)
=
\sum_{\mu=-1}^{+1}(-1)^\mu E^{\mathrm{ext}}_{-\mu}\,\;
b_{Q\mu}^{K,(12)}(I),
\end{equation}
 and
\begin{equation}
\label{b12KQmu}
b_{Q\mu}^{K,(12)}(I)
=
\sum_{g=1,3,5}
G_{fd}^{(g)}\Omega_g^{(K)}
\sum_{\gamma=-g}^{+g}
A^{\mathrm{CF}}_{g\gamma}(I)\,
\cg{1}{\mu}{g}{\gamma}{K}{Q}.
\end{equation}
Note that the superscript $(12)$ indicates that these coefficients arise from the symmetrized second-order perturbative coupling $V_1V_2 + V_2V_1$ between the external electric field ($V_1$) and the crystal-field potential ($V_2$). Explicitly, $B_{KQ}^{(12)}(I)$ and $b_{Q\mu}^{K,(12)}(I)$ are obtained from the contribution of $O_{\mu\gamma}^{1,g}(m_1,m_2)$ in Eq.~(\ref{Om1m2}) (i.e. the $V_1V_2$ term), while the complementary term $O_{\gamma\mu}^{g,1}(m_1,m_2)$ (i.e. the $V_2V_1$ term) yields an identical contribution, resulting in the overall factor of two in Eq.~(\ref{HeffTHIRD}).

Further defining 
\begin{align}
C_{x,KQ}(I) &= \sqrt{2}\,\Big( b_{Q,-1}^{K,(12)}(I) - b_{Q,+1}^{K,(12)}(I) \Big), \\[4pt]
C_{y,KQ}(I) &= i\sqrt{2}\,\Big( b_{Q,-1}^{K,(12)}(I) +  b_{Q,+1}^{K,(12)}(I) \Big), \\[4pt]
C_{z,KQ}(I) &= 2\, b_{Q,0}^{K,(12)}(I).
\end{align}
we can explicitly express the linear dependence of Eq.~(\ref{HeffTHIRD}) on the three Cartesian components of the external electric field $E_\alpha^{\mathrm{ext}}$:
\begin{equation}
\label{HeffFINAL}
	\hat H_{\mathrm{eff}}^{(\mathrm{el})}(I)
	=
	\sum_{\alpha=x,y,z}
	E_\alpha^{\mathrm{ext}}(I)\,
	\hat P_\alpha(I),
\end{equation}
with the on-site effective electric dipole operator for the $I^{th}$ Dy(III) ion:
\begin{equation}
\label{4fElectricDipole}
\hat P_\alpha(I) = 
\sum_{K=2,4,6} \sum_{Q=-K}^{K}
C_{\alpha,KQ}(I) \, \hat U_Q^{(K)}.
\end{equation}
Supplementary Note 6 further shows that the matrix elements of Eq.~(\ref{4fElectricDipole}) on the $|J,M_J\rangle$ basis of each Dy(III) centre, hence for the ab initio crystal field states of interest here, can be straightforwardly evaluated without recurring to coefficients of fractional parentage, in as far as the evaluation is restricted to  a ground multiplet $^6H_{15/2}$  originating from the single Hund's rule $^6H$ Russell-Saunders term.

\paragraph*{Local and Global Magnetoelectric tensors.}

The local magnetoelectric tensor for the $I^{\mathrm{th}}$ Dy(III) ion is obtained by combining the electric Hamiltonian in Eq.~(\ref{HeffFINAL}) with the magnetic Hamiltonian
\begin{equation}
\hat H_{\mathrm{mag}}(I) = - \hat{\mathbf{M}}(I) \cdot \mathbf{B}^{\mathrm{ext,loc}}(I),
\end{equation}
where $\hat{\mathbf{M}}(I) = -\mu_B g_J \hat{\mathbf{J}}(I)$, and applying second-order perturbation theory. This yields
\begin{equation}
\alpha_{\alpha\beta}^{\mathrm{ME}}(I)
=
-
\sum_{n\neq 0}
\left[
\frac{
\langle \mathrm{KD}_0 | \hat P_\alpha(I) | \mathrm{KD}_n \rangle
\langle \mathrm{KD}_n | \hat M_\beta(I) | \mathrm{KD}_0 \rangle
}{E_n - E_0}
+ \mathrm{h.c.}
\right].
\end{equation}
where $\mathrm{h.c.}$ denotes the Hermitian conjugate contribution with
$\hat P_\alpha(I)$ and $\hat M_\beta(I)$ exchanged.

The local tensor is evaluated in a reference frame aligned with the ab initio magnetic axes reported in Tables S1--S5 of Supplementary Note 1. The local ground Kramers doublet $| \mathrm{KD}_0 \rangle$ and excited doublets $| \mathrm{KD}_n \rangle$ are expressed as linear combinations of the local $|J,M_J\rangle$ basis, with coefficients taken from the ab initio wavefunctions of a representative Dy(III) ion and transferred to all sites, consistently with the approximate level of theory adopted. The corresponding wavefunctions are reported in Supplementary Note 6.

The local tensors $\alpha_{\alpha\beta}^{\mathrm{ME}}(I)$ defined above describe the magnetoelectric response associated with the local ground Kramers doublet of the $I^{\mathrm{th}}$ Dy(III) ion. In order to construct the collective magnetoelectric response entering the non-equilibrium toroidal dynamics simulations, we approximate the low-energy collective states of Fe$_{10}$Dy$_{10}$ as Ising-like direct products of local ab initio ground Kramers-doublet states.

For a given collective Ising configuration
\[
\bm{\sigma}=(\sigma_1,\sigma_2,\dots,\sigma_{10}),
\qquad
\sigma_I=\pm,
\]
the corresponding low-energy collective state is approximated as
\[
|0\rangle \equiv |\bm{\sigma}\rangle
=
\prod_{I=1}^{10}
|\sigma_I(\mathrm{KD}_0)\rangle,
\]
where $|\pm(\mathrm{KD}_0)\rangle$ denote the two partners of the local ground Kramers doublet on the $I^{\mathrm{th}}$ Dy(III) ion.

The excited states entering the perturbative magnetoelectric expression Eq.~(\ref{MEtensorMOL}) are correspondingly approximated as local crystal-field excitations on top of the same Ising background,
\[
|n\rangle \equiv |\bm{n}(I,e)\rangle
=
|\sigma_1(\mathrm{KD}_0)\rangle
\cdots
|\sigma_I(\mathrm{KD}_e)\rangle
\cdots
|\sigma_{10}(\mathrm{KD}_0)\rangle,
\]
where $\mathrm{KD}_e$ denotes the $e^{\mathrm{th}}$ excited Kramers doublet of the $I^{\mathrm{th}}$ Dy(III) ion.

Within this approximation, the collective-state-dependent magnetoelectric tensor entering the ultrafast dynamical simulations is written as
\begin{equation}
	\alpha_{\alpha\beta}^{\mathrm{ME}}(\bm{\sigma}) = 
    - \sum_{\bm{n}(I,e)} 
    \left[ 
     \frac
     {\langle \bm{\sigma} | P_\alpha | \bm{n}(I,e) \rangle \langle \bm{n}(I,e) | M_\beta | \bm{\sigma}\rangle}
     {E_{\bm{n}(I,e)} - E_{\bm{\sigma}} }  
     + 
     \frac
     {\langle \bm{\sigma} | M_\beta | \bm{n}(I,e) \rangle \langle  \bm{n}(I,e) | P_\alpha | \bm{\sigma}\rangle}
     {E_{\bm{n}(I,e)} - E_{\bm{\sigma}} } 
     \right].
\end{equation}

Since the local crystal-field excitation energies ($\gtrsim100$ cm$^{-1}$) are much larger than the splittings within the low-energy collective Ising manifold ($\lesssim2.5$ cm$^{-1}$), the denominators are well approximated by the corresponding local crystal-field excitation energies obtained from ab initio calculations. Under these assumptions, the collective tensor reduces to a sum of rotated local contributions weighted by the Ising orientations,
\begin{equation}
\bm{\alpha}^{\mathrm{ME}}(\bm{\sigma})
=
\sum_I
\sigma_I\,
R_I\,
\bm{\alpha}^{\mathrm{ME}}_{\mathrm{loc}}(I)
\,R_I^T,
\end{equation}
where $\sigma_I=\pm1$ labels the orientation of the local ground Kramers doublet on site $I$, and the rotation matrices $R_I$ transform from the local magnetic reference frames to the global molecular frame, and are constructed from the ab initio principal magnetic axes of the $I^{\mathrm{th}}$ Dy(III) ion expressed in the common reference frame of the Fe$_{10}$Dy$_{10}$ structure (Tables S1--S5, Supplementary Note 1).

The magnetoelectric tensor is, in general, non-symmetric with respect to the exchange of its Cartesian indices $\alpha$ and $\beta$. It can therefore be decomposed into a scalar (trace), antisymmetric, and traceless symmetric part:
\begin{equation}
\alpha_{\alpha\beta}
=
\frac{1}{3}\mathrm{Tr}(\alpha)\,\delta_{\alpha\beta}
+
\alpha^{\mathrm{sym,tr}}_{\alpha\beta}
+
\alpha^{\mathrm{anti}}_{\alpha\beta}.
\end{equation}

The antisymmetric component defines an induced magnetoelectric toroidal moment,
\begin{equation}
\tau_\gamma = \epsilon_{\gamma\alpha\beta}\,\alpha^{\mathrm{anti}}_{\alpha\beta}.
\end{equation}

Further implementation details, including projector algebra, tensor recoupling, and explicit evaluation of matrix elements, are provided in Supplementary Note 6.

\section*{Data availability}

The data supporting the findings of this study are available within the paper and its Supplementary Information. Additional numerical data and simulation outputs are available from the corresponding authors upon reasonable request. Source Data are provided with this paper.

\section*{Code availability}
The code required to reproduce the results in this manuscript was developed in-house and is currently under active development. It is available from the authors upon reasonable request.


\section*{Acknowledgements}
The authors thank Jürgen Schnack for valuable discussions.  A. S. acknowledges funding via the grant P-DiSC BIRD2023-UNIPD from the Department of Chemical Sciences of the University of Padova, from the University of Padova and Monash University Joint Initiative in Research (2024 Seed Fund scheme), and from the CINECA award of HPC resources and support under the ISCRA initiative, Project Grant SMTQUANT.
J.B., Y.F.S., C.E.A., W.W. and A.K.P. acknowledge funding by the German Research Council (DFG) through the CRC 1573 “4f for Future”. We furthermore thank the Helmholtz Association for funding through POF MSE.

\section*{Contributions}
A.S. and K.H. developed and implemented the theoretical models and the proposed protocol to generate and observe toroidal polarization in molecular systems.  S.C. performed the DFT calculations.  J.B., Y.F.S., A.B., C.E.A. and A.K.P. performed the synthesis and structural characterization of the Fe$_{10}$Dy$_{10}$ ring. Y.L. performed the SQUID magnetometry measuremnts and W.W. performed MicroSQUID experiments.  M.A. performed the specific heat experiments.  A.S. and K.H. wrote the manuscript with help from J.B., C.E.A., A.K.P. and M.A..  All authors discussed the results, analyzed the data, and contributed to the manuscript.

\bibliographystyle{sn-nature}

\newif\ifarxiv
\arxivtrue 
\ifarxiv
\clearpage
\includepdf[pages=-]{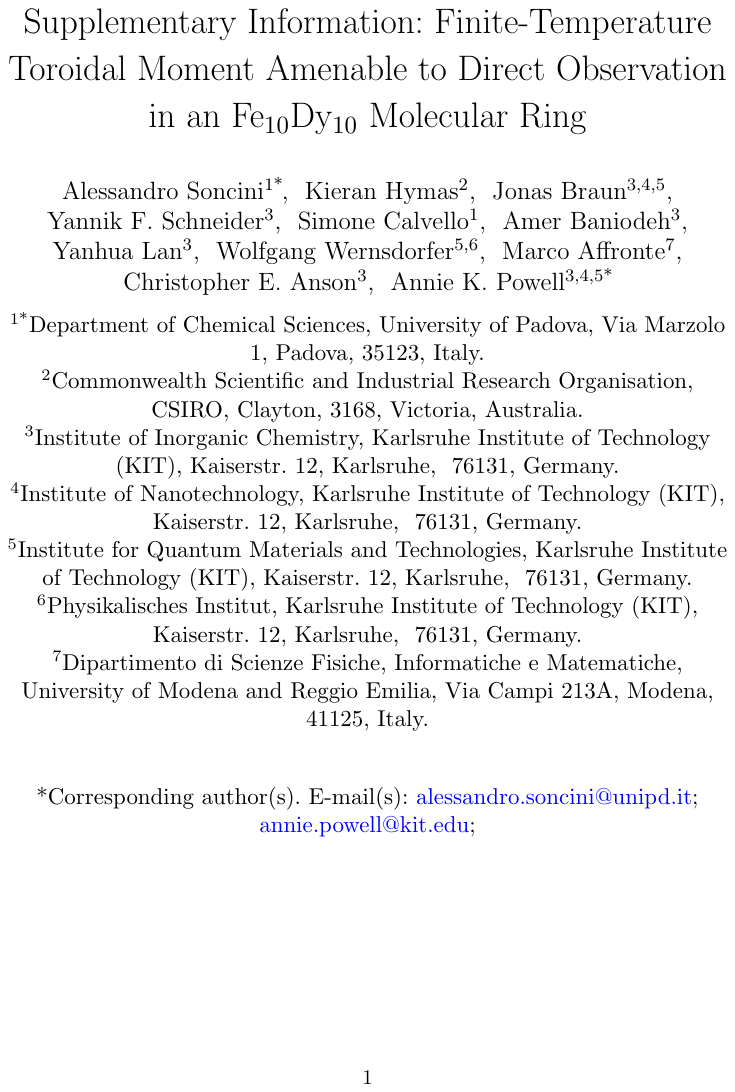 }
\fi

\end{document}